\documentclass[aps,nofootinbib,preprintnumbers,twocolumn,superscriptaddress,prd]{revtex4-1}

\usepackage{amsmath}
\usepackage{amssymb}
\usepackage{slashed}
\usepackage{graphicx}
\usepackage{graphics}
\usepackage{epsfig}
\usepackage{epstopdf}
\usepackage{bbm,bm}
\usepackage{psfrag}
\usepackage{hyperref}
\usepackage{color}
\usepackage{comment}
\usepackage{amsfonts}
\usepackage{dsfont}

\graphicspath{{figurespdf/}}

\newcommand{\be}{\begin{eqnarray}}
\newcommand{\ee}{\end{eqnarray}}
\newcommand{\Eqref}[1]{Eq.~\eqref{#1}}

\begin{document}

\title{Momenta fields and the derivative expansion}


\author{Luca Zambelli}
\email{luca.zambelli@uni-jena.de}
\affiliation{\mbox{\it Theoretisch-Physikalisches Institut, Friedrich-Schiller-Universit{\"a}t Jena,}
\mbox{\it D-07743 Jena, Germany}}



\begin{abstract} 

The Polchinski exact renormalization group equation 
for a scalar field theory in arbitrary dimensions
is translated, by means of a covariant Hamiltonian formalism, into a
partial differential equation for an effective Hamiltonian density that depends 
on an infinite tower of momenta fields with higher spin. 
A natural approximation scheme is then expanding the Hamiltonian in momenta with increasing rank.
The first order of this expansion, one next to the local potential approximation, is 
regulator-independent and
 already includes infinitely many derivative interactions.
Further truncating this 
down to a quadratic dependence on the momenta leads to an alternative to the first order of the derivative expansion,
which is used to compute $\eta=0.03616(1)$ for the critical exponent of the three dimensional Ising model.
\end{abstract}


\maketitle

\section{Introduction and Summary}
\label{sec:intro}

This work develops a new method to construct approximate solutions 
of functional renormalization group (FRG) equations.
The latter provide an exact representation of quantum or statistical field theory~\cite{Wilson:1973jj},
and as such they can be used to describe, at least in principle, all kinds of nonperturbative phenomena.
Whether this is feasible in practice solely depends on being able to truncate the FRG equations 
to a solvable set of equations that encode the wanted pieces of information~\cite{ReviewRG}.
Though it is desirable to adjust approximations to each specific problem,
these should allow for some control, in the form of progressive improvement,
such that it is necessary to formulate systematic strategies, and not simply {\it ad hoc} ans\"atze.
Furthermore, since any truncation induces errors that can be hard to estimate and reduce, 
it is of primary importance that a rich pool of approximation schemes be available. 
We therefore believe that developing one more of them might be a worthy endeavor,
even after forty years of FRG studies.

While dealing with infinitely many interaction terms is not a problem, 
as shown by the successes of the local potential approximations (LPA),
which include generic functions of constant fields~\cite{Nicoll:1974zz}, 
the description of nontrivial momentum dependence of correlation functions 
is less straightforward and triggered the development of several approximation strategies.
Historically two such schemes have been playing a prominent role in the scientific community.
The first is the vertex expansion (VE), which is an expansion in field variables while retaining the full momentum dependence~\cite{Ellwanger:1993mw,Morris:1993qb}.
The infinite tower of flow equations for the $n$-point functions is therefore truncated by
considering only vertices up to a given number of legs.
The second is the derivative expansion (DE), that is an expansion in powers of momenta while retaining the full field dependence~\cite{Golner:1985fg}.
This approximation makes use of local actions
with definite given powers of field derivatives, and it relies on the assumption that the system possesses at least one mass-scale $m$,
below which higher powers of $(-\partial^2/m^2)$ play a progressively less important role.
As a matter of fact, however, it 
performs quite well also in describing conformal models and critical 
phenomena~\cite{Golner:1985fg,Tetradis:1993ts,Ball:1994ji,Morris:1996kn,Morris:1997xj,Canet:2003qd,Litim:2010tt},
that since the early history of the FRG have been a standard application and 
a reference point for improvements of approximation methods.
For specific examples of applications of these two approximation schemes we refer the reader to
the reviews~\cite{ReviewRG}.

As it is to be expected, several alternatives to these two complementary approaches
have been developed.
The method introduced by Blaizot, Mendez-Galain and Wschebor (BMW)~\cite{Blaizot:2005xy,Benitez:2011xx}
 allows one to keep infinitely many vertices  
into account as in the DE, while retaining generic functions of momenta
as in the VE. This is accomplished by making use of the flow equations for the $n$-point functions, 
but neglecting part of the momentum dependence of some vertices.
Much closer to the DE is the scaling fields expansion (SFE)~\cite{Wegner1976,Riedel:1986re,Newman:1984zz},
that is an expansion in eigenoperators of the linearized flow around the 
Gau\ss ian FP.
This method has been recently revived in~\cite{O'Dwyer:2007ia}, where it was critically
compared to the DE. 
By making use of several features of the SFE, the same authors proposed
to apply a derivative expansion to the action expressed in terms 
of a normal ordered basis of monomials of $\phi$~\cite{Osborn:2011kw}. 
In the following we will refer to this approach by the name of normal ordered derivative expansion (NDE).
Other methods have been proposed, for instance~\cite{Golner:1998sr} and~\cite{Hasselmann:2012xw},
but we do not aim at a comprehensive enumeration.

The construction of systematic approximations goes together with the desire to minimize
problems, and not just to maximize the accessible amount of information.
In fact, apart for the sometimes unknown effect of a truncation of the theory space, 
the FRG results are affected by other ambiguities, which can be considered as consequences of the former. 
Among these is the possible breaking of some wanted symmetries.
A simple example is the ubiquitous symmetry under linear rescalings of the fields, 
which is sometimes referred to as linear reparametrization invariance.
This is generically broken by the above mentioned
truncation schemes, but it can be restored by an appropriate choice of regulator 
functions~\cite{Ball:1994ji,Morris:1996kn}.
Indeed, another truncation-induced ambiguity is the dependence on the regularization itself,
which often affects quantities that should be universal.
In fact, regulator dependence is present at any order of the DE beyond the LPA,
as well as in the VE, in the SFE, and in the BMW method, though some approximations
might result in a less severe dependence than others.
Clearly, the regulator choice is not totally free and it is actually possible to optimize it for
each approximation~\cite{Litim:2001up}.
We do not know if it is possible to devise a systematic truncation strategy that preserves regularization independence
of universal quantities and reparametrization invariance of the exact RG equations at any order.
The goal of this work is a much easier one, yet such issues will be relevant in the present discussion.

By taking a conservative point of view, we would like
to facilitate the computation of high orders of the DE.
This will result in the construction of an approximation scheme that is not the DE,
though it is extremely close to it.
While the DE has been already pushed to the second order ($\partial^4$)~\cite{Canet:2003qd,Litim:2010tt}
and to the third order ($\partial^6$)~\cite{Delamotte:ERG2014}
for a real scalar field theory, it is nowadays hard to imagine to go 
very far beyond these orders.
The combinatoric computational difficulty in obtaining the flow
equations for a high order of the DE is essentially a particular
form of the combinatoric difficulty of dealing with high orders of a Taylor expansion,
the variable of expansion being momentum $p^2$.
The standard way of circumventing this combinatoric problem is to compute the flow of a full
function at once, which would require to keep a generic off-shell value for $p^2$.
Yet, in a DE setup one would also need to keep $\phi$ generic and constant, in order 
to describe infinitely many vertices.
The traditional way to achieve these conditions simultaneously is by means of a Hamiltonian formalism,
whose application to the FRG is the subject of the present work.

In order to implement this idea 
we confine our discussion to a real $\mathbb{Z}_2$-symmetric
scalar field theory,
and we restrict ourselves to a specific class of
truncations which can be essentially identified with an arbitrarily high order of the DE.
The RG flow of these truncations will be analyzed in Sec.~\ref{sec:Lagrangian},
for the specific case of the Polchinski equation~\cite{Polchinski:1983gv},
where we discuss how different regularizations, or coarse-graining
prescriptions, lead to structurally different truncated equations.
In Sec.~\ref{sec:Hamiltonian} we apply the above mentioned idea by
describing the Hamiltonian translation of these equations,
which amounts to replacing the arbitrary-order derivative
\be
\frac{\mathrm{d}\phantom{LL} }{\mathrm{d}x^{\mu_m}} \dots \, \frac{\mathrm{d}\phantom{LL} }{\mathrm{d}x^{\mu_1}}\phi(x)
\longrightarrow \pi^{\mu_1\dots \mu_m} \nonumber
\ee
with a symmetric tensor field.
An important conceptual feature of this program is 
that derivative terms with different tensorial signature are mapped into different 
momenta structures.
Thus, two terms like $\phi(-\partial^2)\phi$ and $(\partial\phi)^2$ get
translated into two very different objects, 
$\phi \pi^{\mu\nu}\delta_{\mu\nu}$ and $\pi^\mu\pi_\mu$ respectively.
These can be related only through canonical transformations.
The systematic approximation scheme we discuss in this work
breaks the latter invariance by treating these and other equivalent terms on different footing,
and therefore significantly departs from the DE.
In fact, we are interested in an expansion of the effective Hamiltonian 
in momenta fields, organized by including momenta in increasing-rank order.
More precisely, only the dependence of the Hamiltonian on rank-one momenta
will be explicitly addressed in this work.
The corresponding truncated flow equations are derived in  Sec.~\ref{sec:Hamiltonian}
and describe infinitely many derivative interactions in the form of an arbitrary function of $\pi^\mu$.
One of these equations, descending from a convenient regularization scheme,  
takes the simple form of a second order partial differential equation for a function of $\phi$ and $\pi^\mu$.
Like the LPA, this is regulator independent, thanks to the freedom to rescale the additional
variable $\pi^\mu$.

In Sec.~\ref{sec:critical} we discuss a first application of this approximation, 
with the only purpose to check for conceptual mistakes in the derivation of these equations
and to assess whether this different treatment of the derivative sector of the effective action 
properly captures nonperturbative effects and universal phenomena.
We conservatively turn to the study of the Wilson-Fisher fixed point (FP) in three dimensions,
in a simplified set up where we neglect all the interactions that are more than quadratic in the momenta,
in order to make contact with the rich literature about the first order of the DE.
Since we do not aim at a comprehensive study but only at a first exploration of 
the properties of the truncations proposed here, we focus our attention on the computation of the 
critical exponent which is the most sensitive to the derivative sector, 
as well as the most challenging to accurately estimate by means of the FRG,
namely the critical anomalous dimension $\eta$.

In Tab.~\ref{tab:comparing_eta} we compare our result to some literature, especially from the state-of-the-art FRG computations.
Some details are given about the specific implementations of the FRG:
 first of all which exact equation is used (Wilson~\cite{Wilson:1973jj}, Wetterich~\cite{Wetterich:1992yh,Morris:1993qb}, Polchinski~\cite{Polchinski:1983gv}),
then the truncation scheme and its order. With `bf' we specify the use of the background field method,
while `implicit' refers to the use of implicit optimization.
For more details and a collection and comparison of FRG predictions for $\eta$ at various orders of the DE we refer
the reader to~\cite{Litim:2010tt}. 
About the NDE reference, it should be stressed that the equations in~\cite{Osborn:2009vs}
are obtained from the order ($\partial^2$) of the NDE by neglecting a specific term. 
This allows to analytically perform a choice of regulator-dependent coefficients such that linear reparametrization invariance is satisfied.
We also recall some results from high-temperature expansions~\cite{Campostrini:2002cf},
Monte-Carlo methods~\cite{Hasenbusch:2011yya} and the conformal bootstrap~\cite{Simmons-Duffin:2015qma}.

Regarding our estimate, the uncertainty is numerical, as it is explained in Sec.~\ref{sec:critical}, 
hence it is possible to further reduce it with a more accurate analysis.
Yet, this must also be interpreted as an uncertainty on the uniqueness of this result
against the change of one arbitrary parameter, namely the overall normalization of the FP Hamiltonian.
We did not observe any dependence of $\eta$ on such parameter,
which is compatible with linear reparametrization invariance. 
However we have no proof for the exactness of this invariance,
since we only observed it within the limitations of a numerical sampling.
Through a study of the linearized equations around the FP it will
be possible to extract more critical exponents and to further test the presence of an exact marginal perturbation,
which would make the results of this truncation completely unique, i.e. regulator and parametrization invariant.
We postpone this analysis to future works.
However, let us remark that for the determination of the relevant critical exponent $\nu$ the present 
treatment is expected to change (lower) the LPA estimate $\nu=0.65$ only by few percents, 
as it happens in the DE where at first order ($\partial^2$) this becomes $\nu=0.62\div0.63$~\cite{Litim:2010tt}.

\begin{table}
\resizebox{8cm}{1.9cm}{
\begin{tabular}{|l|l|l|l|l|l|l|l|l|l|l|l|l|}
\hline
ref 							& year			 & method				& info 				 				&$\eta$	\\
\hline
\cite{Newman:1984zz}		&1984			& FRG			& Wil, SFE $O(n_\text{eq}\!=\!10)$			&0.040(7)	\\
\cite{Canet:2003qd}			&2003			& FRG			& Wet, DE $O(\partial^4)$ 					&0.033		\\ 
\cite{Osborn:2009vs} 		&2009 			& FRG			& Pol, NDE $O(\partial^2)$					&0.041347	\\
\cite{Litim:2010tt}			&2010			& FRG			& Wet, DE $O(\partial^4)$ bf					&0.0313		\\ 
\cite{Litim:2010tt}			&2010			& FRG			& Wet, DE $O(\partial^4)$ bf, implicit			&0.034		\\ 
\cite{Benitez:2011xx}			&2011			& FRG			& Wet, BMW $O(s=2)$ 						&0.039		\\
\cite{Campostrini:2002cf}		&2002			&HT				&O(25)										&0.03639(15)\\
\cite{Hasenbusch:2011yya}	&2011			&Monte-Carlo	&											&0.03627(10)\\
\cite{Simmons-Duffin:2015qma}&2015			&CB				&											&0.036302(12) \\
							&				&FRG			& this work									&0.03616(1)	\\ 
\hline
\end{tabular}
}
\caption{The critical exponent $\eta$ for the three-dimensional Ising universality class, from the functional renormalization group,
high-temperature expansions, Monte-Carlo simulations and the conformal bootstrap. For the meaning of the abbreviations see the main text.}
\label{tab:comparing_eta}
\end{table}

To conclude, few comments are in order regarding the relation of this work with 
other lines of research that might look close to the present project.
 There have been several works on relating exact RG equations to higher-spin theories, that also
translate the arbitrarily high order momentum dependence of the effective action into a dependence
of the action on higher-spin variables~\cite{Akhmedov:2010sw}. 
Though these basic facts are a starting point for these studies as well as for the present one, 
the goals and the methods are very different, and our discussion does not add any contribution in those directions.
On the other hand, these attempts to reconstruct higher-spin theories and corresponding
gauge symmetries from the full nonlinear exact RG equations might shed light on the 
role of canonical transformations and on the structure of higher orders of the truncations
addressed here.

Certainly related to this work is the analysis of FRG equations of effective Hamiltonian actions
outlined in~\cite{Vacca:2012vt}. Though partly motivating this piece of research,
the goal of that formalism is the nonperturbative study of those particular field theories
that, in a first order formulation, show an interacting momentum sector.
Examples of such models are for instance those with bare Hamiltonians more than quadratic in momenta or 
those that possess nonlinear symmetries.
In these cases, the DE of such Hamiltonian actions differ significantly from the DE of corresponding Lagrangian actions,
and functions of momenta would effectively incorporate infinitely many derivatives of configuration variables.
However, for ordinary theories with quadratic bare Hamiltonians such a formalism would boil down to the usual 
Lagrangian FRG.
This is not so for the present construction, that allows for a novel treatment of almost any model.
Also, the effective Hamiltonian action discussed in~\cite{Vacca:2012vt} is still a generic functional
that can depend on derivatives of $\phi$ and $\pi$, since the Hamiltonian translation is
performed at the level of the classical theory, and each RG step produces again non-localities and
higher derivatives.
What the present formulation does, instead, is implementing a Hamiltonian translation after each RG
step, thus keeping a derivative-free effective Hamiltonian all along the RG flow.

\section{RG equations for local effective Lagrangians}
\label{sec:Lagrangian}

An operative definition of field theory needs a regularization introducing a scale
$\Lambda$. Yet, observables should be to some extent independent of such a
regularization.
This is made possible by the dependence of the microscopic dynamics,
as encoded for example in the Wilson effective action $S$, on the scale itself.
Thus, changing the value of $\Lambda$, must result in a change of $S$.
The standard way of interpreting such a variation is identifying it with a nonlinear redefinition of the fields,
such as a change of variables in the path integral~\cite{Latorre:2000qc}, which therefore ensures 
the invariance of observables.
Different choices of field redefinitions lead to different RG equations for $S$.
In this work we will concentrate on the Polchinski equation, that reads
\be
\dot{S}[\phi]=\frac{1}{2}\int_{x y}\frac{\delta S[\phi]}{\delta\phi(x)}\dot{C}(x-y)\frac{\delta S[\phi]}{\delta\phi(y)}\nonumber\\
-\frac{1}{2}\int_{x y}\frac{\delta\phantom{S}}{\delta\phi(x)}\dot{C}(x-y)\frac{\delta S[\phi]}{\delta\phi(y)}
\ee
where dotted quantities are differentiated with respect to $t=-\log\Lambda$.
The meaning of $\dot{C}(x-y)$ and the constraints on it are traditionally understood in terms of its Fourier transform
$\dot{\hat{C}}(p^2)$.
Then, one can think about $\dot{\hat C}$ as the rate of change $-\Lambda\partial_\Lambda \hat{C}$ of some
regularized propagator
\be
\hat{C}(p^2)=\Lambda^{2d_\phi -d}\ \ \frac{\Lambda^2}{p^2}K\!\left(\frac{p^2}{\Lambda^2}\right)\ .
\ee
Here $d_\phi$ is the full (quantum) dimensionality of the field $\phi$ and 
$K$ is a cutoff function that is intended to regulate the UV and/or IR behavior of this propagator.
Sticking to traditional notations, we will split the quantum dimensionality of $\phi$ into a canonical part and an anomalous one
$d_\phi=(d-2+\eta)/2$.
In this work we will take $\eta$ as independent of $\Lambda$, since we will later address the case of theories at fixed points of the RG.

We are now interested in local truncations, which correspond to a Lagrangian density depending on generically high derivatives of the fields.
For the sake of notational simplicity,
let us introduce multi-indices $M\equiv(\mu_1, ... , \mu_m)$ with $m\in\mathbb{N}$ and denote
\be
\phi_M=\phi_{\mu_1 ... \mu_m}(x)=\frac{\mathrm{d}\phantom{LL} }{\mathrm{d}x^{\mu_m}} ...\, \frac{\mathrm{d}\phantom{LL} }{\mathrm{d}x^{\mu_1}}\phi(x)
=\frac{\mathrm{d}\phantom{LL} }{\mathrm{d}x^M}\phi(x) \ . \ \ 
\ee
We will use a similar multi-index notation for derivatives of any other function.
In formulas, we address the following truncation
\be\label{eq:local_effective_Lagrangian}
S[\phi]=\int_x {\cal L}(x,\phi_M(x))\ .
\ee	
Let us remark that also some nonlocal actions can be rewritten in terms of Lagrangian densities of the present kind, possibly depending explicitly on the position in space, by just expanding all fields in a Taylor series around a common point $x$. 
The goal of this section is rewriting such a truncation of the Polchinski equation as a partial differential equation for $\cal L$.
To this end we need to express the functional derivative of the action as a standard Euler-Lagrange operator acting on the Lagrangian density
\be
\frac{\delta S}{\delta\phi(x)}&\!=\!&(-)^M\frac{\mathrm{d}\phantom{LL} }{\mathrm{d}x^M}\frac{\partial{\cal L}}{\partial\phi_M}(x)\\
&\!=\!& \frac{\partial{\cal L}}{\partial\phi}-\frac{\mathrm{d}\phantom{LL} }{\mathrm{d}x^{\mu_1}}\frac{\partial{\cal L}}{\partial\phi_{\mu_1}}+
\frac{\mathrm{d}^2\phantom{LL} }{\mathrm{d}x^{\mu_1}\mathrm{d}x^{\mu_2}}\frac{\partial{\cal L}}{\partial\phi_{\mu_1\mu_2}}+ ...\nonumber
\ee
where we used, here and in the rest of the paper, Einstein's summation convention for multi-indices, and we denoted $(-)^M=(-1)^m$
for a multi-index of length $m$. We will write $M$ in place of its length $m$ also in other similarly situations,
when this appears as an unambiguous abuse of notations.
Notice that in the sum above we included the empty index with $m=0$, corresponding to $\phi_M(x)=\phi(x)$,
which sometimes we will refer to by simply writing $M=0$.
The second order functional derivative can also be rewritten in terms of infinitely many partial derivatives.
First of all, it is convenient to rewrite the first order derivative as the integral of a Lagrangian density
\be
\frac{\delta S}{\delta \phi(x)}
=\int_y \frac{\partial{\cal L}}{\partial\phi_M}(y) \delta_{M}(y-x) =\int_y {\cal L}^{(1)}_x(y) \ .
\ee
Then, one can iterate the application of the Euler-Lagrange operator
to get the first order functional derivative of an action with Lagrangian ${\cal L}^{(1)}_x$
\be
\frac{\delta^2 S}{\delta\phi(y)\delta\phi(x)}
=(-)^N\frac{\mathrm{d}\phantom{LL} }{\mathrm{d} y^N}\left[\frac{\partial^2{\cal L}}{\partial\phi_N\partial\phi_M}(y)\delta_{M}(y-x)\right] .\ 
\ee
Applying this formula to the quantum term in the flow equation, integrating by parts and dropping the integrals of total derivatives, this can be written as
\be
\int_{x y}\!\!\dot C (x-y)\frac{\delta^2 S}{\delta\phi(y)\delta\phi(x)}
=(-)^N \dot C_{MN}(0) \!\int_x \frac{\partial^2 {\cal L}}{\partial\phi_N\partial\phi_M}(x)\nonumber 
\ee
where we assumed that the regularized propagator is an even function of the position in space $C(x)=C(-x)$.
As a consequence, one can recast the Polchinski equation for the present truncation in the form
\be\label{eq:flowLnonlocal}
\int_x \dot{\cal L}(x)&=&\frac{(-)^N}{2}\left\{\int_{x y}\frac{\partial{\cal L}}{\partial\phi_M}(x)\dot{C}_{MN}(x-y)\frac{\partial{\cal L}}{\partial\phi_N}(y)
\right.\nonumber\\
& &-\left.\dot{C}_{MN}(0)\int_x \frac{\partial^2{\cal L}}{\partial\phi_M\partial\phi_N}(x)\right\} .
\ee
This clearly shows that, despite our initial ansatz for the effective action in~\Eqref{eq:local_effective_Lagrangian} was assuming a local effective Lagrangian,
the right hand side (r.h.s.) of the flow equation generates nonlocalities through the classical term.
Projecting these nonlocalities out of our truncation would be too crude an approximation,
and one would miss a crucial part of the interplay between pointlike interactions and the derivative 
sector~\footnote{I am grateful to T. R. Morris for pointing this out to me, and for suggesting
the different treatment of this term that is described in what follows.}.
We then follow the same principle inspiring the presence of infinitely many derivatives in
\Eqref{eq:local_effective_Lagrangian}, that at least part of this nonlocal structure could be
rewritten as a higher-derivative local dynamics.
In order to project the r.h.s. of the Polchinski equation onto a local effective Lagrangian, we then expand the integrand of the classical term about a 
single point
\be\label{eq:Taylor_yx}
\frac{\partial{\cal L}}{\partial\phi_N}(y)=
\frac{\partial{\cal L}}{\partial\phi_N}(x)+
\frac{(y-x)^L}{L!}\frac{\mathrm{d}\phantom{Ll} }{\mathrm{d}x^L}\frac{\partial{\cal L}}{\partial\phi_N}(x)
\ee
where $L\neq0$.
Since ${\cal L}$ depends on $\phi_M$ and possibly also separately on $x$, we can further
split
\be
\frac{\mathrm{d}\phantom{Ll} }{\mathrm{d}x^\lambda}=\frac{\partial\phantom{L}}{\partial x^\lambda}+\phi_{M\lambda}\frac{\partial\phantom{L}}{\partial\phi_M}
\ee
where $\partial_\lambda$ denotes the $x^\lambda$-derivative at fixed $\phi_M$.
Such a contribution is nonvanishing only if the couplings in $\cal H$
 have an explicit space, i.e. momentum, dependence.
In this paper we will restrict ourselves to the pointlike interaction limit and neglect this explicit $x$-dependence.
The $L-$th derivative can then be written as
\be
\frac{\mathrm{d}\phantom{F} }{\mathrm{d}x^L}=
\sum_{i=1}^{L}\phi_{(M_1}\dots\phi_{M_i)_L}\frac{\partial^i \phantom{F}}{\partial \phi_{M_1}\dots \partial \phi_{M_i}}
\ee
where $\phi_{(M_1}\dots\phi_{M_i)_L}$ denotes a sum over all possible  ways of distributing the indices inside $L$
on the $i$ entries $\phi_{M_1}\dots \phi_{M_i}$, under the rules that ordering inside each entry does not matter,
 that permutations of $M_1\dots M_i$ do not matter,
and that there must be at least one index out of $L$ per entry.
For instance
\be
\frac{\mathrm{d^2}\phantom{F} }{\mathrm{d}x^{\lambda_1}\mathrm{d}x^{\lambda_2}}\!=\!
\phi_{M_1{\lambda_1}{\lambda_2}}\frac{\partial \phantom{F}}{\partial \phi_{M_1}}\!+
\phi_{M_1{\lambda_1}}\phi_{M_2{\lambda_2}}\frac{\partial^2 \phantom{F}}{\partial \phi_{M_1}\partial\phi_{M_2}}
\nonumber \ .
\ee
This gives rise to a factorization of regulator-dependent coefficients as in the quantum term,
of the form
\be
 J_{L,MN}=\int_y y^L \dot{C}_{MN}(y)\ .
\ee
Whenever $M+N>L$, integrating by parts and assuming that the regulator is such that the boundary terms vanish, one gets $J_{L,M+N}=0$.
For instance, for $L=0$  the coefficient $J_{0,M+N}$ 
is the integral of a total derivative and vanishes unless $M=N=0$,
in which case $J_{0,0}=\dot{\hat{C}}_\Lambda(0)$.
To sum up, the projection of the Polchinski equation on the ansatz of 
a local effective Lagrangian gives (the spatial integral of) the following flow equation
\be\label{eq:flowLcomplete}
&\dot{\cal L}&=
\frac{1}{2}\dot{\hat{C}}(0) \left(\frac{\partial{\cal L}}{\partial\phi}\right)^2 -\frac{(-)^N}{2}\dot{C}_{MN}(0)
\frac{\partial^2{\cal L}}{\partial\phi_M\partial\phi_N}\\
&+& \frac{(-)^M J_{L,MN}}{2L!}\frac{\partial{\cal L}}{\partial\phi_M}
\phi_{(M_1}\dots\phi_{M_i)_L}\frac{\partial^{i+1} {\cal L}}{\partial \phi_{M_1}\dots \partial \phi_{M_i}\partial\phi_N}\nonumber
\ee
where $L\neq0$, while the sum over $M_i, M, N$ includes the empty index, and for 
notational simplicity we dropped the summation
symbol for $i=1\dots L$.
The first term comes from the zeroth order $(y-x)^0$ of the Taylor expansion of the classical term,
for which we assumed that only the $M=N=0$ contribution is nonvanishing. This is because we
do not want IR divergences and thus require $\dot{\hat{C}}(0)<\infty$, which entails that any
integral of a total derivative, such as $\int_y \dot{\hat{C}}_M (y)$ for $M>0$, vanishes.

The stucture of \Eqref{eq:flowLnonlocal} has been considerably complicated by the procedure of Taylor
expansion, that involves a sum of infinitely many terms.
One may wonder if
upon truncation of this equation by neglecting the dependence
on derivatives of order bigger then some integer $k$,
 i.e. the dependence of ${\cal L}$ on $\phi_M$ whenever $M>k$,
 this sum gets finite.
This is not the case already for $k=1$.
In fact, one is forced to set $M_1=\dots=M_i=0$ but still the sum over $i$ remains
\be\label{eq:flowL}
\dot{\cal L}&=&
\frac{1}{2}\dot{\hat{C}}(0) \left(\frac{\partial{\cal L}}{\partial\phi}\right)^2
 -\frac{1}{2}\dot{C}(0)\frac{\partial^2{\cal L}}{\partial\phi^2}+\frac{1}{2}\dot{C}_{\mu\nu}(0)
\frac{\partial^2{\cal L}}{\partial\phi_\mu\partial\phi_\nu}\nonumber\\
+\!\!&&\!\!\!\!\frac{J_{\lambda_1\cdots\lambda_i,\mu}}{2i!}
\phi_{\lambda_1}\dots\phi_{\lambda_i}
\left(\frac{\partial{\cal L}}{\partial\phi}\frac{\partial^{i+1}{\cal L}}{\partial \phi^i\partial\phi_\mu}
-\frac{\partial{\cal L}}{\partial\phi_\mu}\frac{\partial^{i+1} {\cal L}}{\partial \phi^{i+1}}\nonumber\right)\\
-\!\!&&\!\!\!\!\frac{J_{\lambda_1\cdots\lambda_i,\mu\nu}}{2i!}
\phi_{\lambda_1}\dots\phi_{\lambda_i}
\frac{\partial{\cal L}}{\partial\phi_\mu}\frac{\partial^{i+1} {\cal L}}{\partial \phi^i\partial\phi_\nu}\nonumber\\
-\!\!&&\!\!\!\!\frac{J_{\lambda_1\cdots\lambda_i,0}}{2i!}
\phi_{\lambda_1}\dots\phi_{\lambda_i}
\frac{\partial{\cal L}}{\partial\phi}\frac{\partial^{i+1} {\cal L}}{\partial \phi^{i+1}}\ .
\ee
This is no longer the case if one further projects the flow on the sector quadratic in $\phi_\mu$,
which selects the following terms
\be\label{eq:flowLquadratic}
\dot{\cal L}&=&
\frac{1}{2}\dot{\hat{C}}(0) \left(\frac{\partial{\cal L}}{\partial\phi}\right)^2
 -\frac{1}{2}\dot{C}(0)\frac{\partial^2{\cal L}}{\partial\phi^2}+\frac{1}{2}\dot{C}_{\mu\nu}(0)
\frac{\partial^2{\cal L}}{\partial\phi_\mu\partial\phi_\nu}\nonumber\\
\!&+&\!\frac{J_{\lambda,\mu}}{2}\phi_{\lambda}
\left(\frac{\partial{\cal L}}{\partial\phi}\frac{\partial^{2} {\cal L}}{\partial \phi\partial\phi_\mu}
-\frac{\partial{\cal L}}{\partial\phi_\mu}\frac{\partial^{2} {\cal L}}{\partial \phi^2}\right)\nonumber\\
\!&-&\!\frac{J_{\lambda\mu,0}}{4}\phi_{\mu}\phi_{\lambda}
\frac{\partial{\cal L}}{\partial\phi}\frac{\partial^{3} {\cal L}}{\partial\phi^3} \ .
\ee
Notice that the two terms inside the bracket in the second line are equal
since they differ by a total derivative.
By the same reasoning the last term can be rewritten as
\be
+\frac{J_{\lambda\mu,0}}{4}\phi_{\mu}\phi_{\lambda}\left(\frac{\partial^{2} {\cal L}}{\partial\phi^2}\right)^2\ .
\ee
Later on we will change notation for the regulator-dependent terms in \Eqref{eq:flowLquadratic},
adopting the conventions of~\cite{Ball:1994ji} 
for the following positive quantities
\be
-\frac{1}{2}\dot{\hat{C}}(0)&=&\Lambda^{\eta-2} K_0\nonumber\\
\frac{J_{\lambda,\mu}}{2}&=&-\frac{1}{2}\delta_{\lambda\mu}\dot{\hat{C}}(0)=\delta_{\lambda\mu}\Lambda^{\eta-2} K_0\nonumber\\
-\frac{1}{4}J_{\lambda\mu,0}&=&\frac{\delta_{\lambda\mu}}{4d}\left[
\frac{\partial^2\dot{\hat{C}}}{\partial p_\nu\partial p^\nu}\right]_{p=0}\!\!=\delta_{\lambda\mu}\Lambda^{\eta-4}K_1\nonumber\\
-\frac{1}{2}\dot{C}(0)&=&\Lambda^{d-2+\eta}I_0\nonumber\\
\frac{1}{2}\dot{C}_{\mu\nu}(0)&=&\delta_{\mu\nu}\Lambda^{d+\eta}\frac{I_1}{d}\ .
\ee

In obtaining \Eqref{eq:flowLcomplete} it was crucial to assume that the regulator $C$ acts as a kernel in position or
Fourier space. Specifically, we assumed that $C$ is a differential operator based on a total derivative in position space,
such as 
\be\label{eq:operatorC}
\dot{C}(x-y)=\dot{\hat{C}}\left(-\frac{\mathrm{d}^2\phantom{Ll}}{\mathrm{d} x^\mu \mathrm{d} x_\mu}\right)\delta(x-y)\ .
\ee
Expanding $\dot{\hat{C}}$ in series around zero and evaluating the total derivatives,
leads to the infinite sum in \Eqref{eq:flowLcomplete}, that we previously 
derived by the equivalent procedure of replacing a function of point $y$ by its Taylor series 
in powers of $(y-x)$.
In the present context, where ${\cal L}$ is treated as a function of infinitely many independent
variables $\{\phi, \phi_\mu, \phi_{\mu\nu},\dots ,\phi_M, \dots\}$, this appears to be one
among several other possible choices, corresponding to the freedom to keep some of the
$\phi_M$'s constant while taking the spatial derivatives.
This possibility, that could be unapparent in a Lagrangian formulation, will take a more familiar
shape in the Hamiltonian formulation of the next section.
Since we are interested in studying truncations inspired by the derivative expansion, 
it would be helpful to define RG transformations that are as simple as possible in the
low-momenta sectors. Hence, for definiteness, let us discuss the possibility that the
undifferentiated field itself is kept constant by the coarse-graining operator $\dot{\hat{C}}$.
Again this can be formalized in two equivalent ways. The first one is the procedure of Taylor expansion
in powers of $(y-x)$, in which the undifferentiated field must be considered $y$-independent,
such that the derivatives that in \Eqref{eq:Taylor_yx} were total, now do not hit $\phi$,
\be\label{eq:derivative_constant_phi}
\left.\frac{\mathrm{d}\phantom{L} }{\mathrm{d}x^\lambda}\right|_{\phi}=\partial_\lambda+\phi_{\mu\lambda}\frac{\partial\phantom{L}}{\partial\phi_\mu}
+\phi_{\mu\nu\lambda}\frac{\partial\phantom{L}}{\partial\phi_{\mu\nu}}+\dots\ .
\ee
Then clearly the Taylor expansion of the nonlocality in the classical term will produce corrections that,
with the only exception of the zeroth order term,
do not affect a truncation projecting on the sector of $\phi$ and $\phi_\mu$.
As a consequence, in this case one simply needs to take the two functional derivatives
in the classical term
\be
\frac{(-)^{M+N}}{2}\!\int_{x y}\!\dot{C}(x-y)
\frac{\mathrm{d}\phantom{Ll} }{\mathrm{d}x^M}\frac{\partial{\cal L}}{\partial\phi_M}(x)
\frac{\mathrm{d}\phantom{Ll} }{\mathrm{d}y^N}\frac{\partial{\cal L}}{\partial\phi_N}(y)
\ee
and evaluate them at the same space-point.
The remaining $y$-integral factorizes a $\dot{\hat{C}}(0)$.

The second equivalent way of reaching this conclusion is directly replacing 
$\dot{C}(x-y)$ in the last equation with the expression in \Eqref{eq:operatorC},
where the Laplacian now is to be understood as a differentiation at constant $\phi$
as in \Eqref{eq:derivative_constant_phi},
and then expanding $\dot{\hat{C}}$ around zero argument.
Since the Laplacian at fixed $\phi$ will produce only terms with momenta of rank bigger than one,
only the zeroth order term will contribute to the selected truncation.
Hence, in both cases, one obtains a simpler alternative to \Eqref{eq:flowL}, namely
\be\label{eq:flowLsimpler}
\dot{\cal L}&=&
\frac{1}{2}\dot{\hat{C}}(0) \left(\frac{\partial{\cal L}}{\partial\phi}\right)^2
 -\frac{1}{2}\dot{C}(0)\frac{\partial^2{\cal L}}{\partial\phi^2}+\frac{1}{2}\dot{C}_{\mu\nu}(0)
\frac{\partial^2{\cal L}}{\partial\phi_\mu\partial\phi_\nu}\nonumber\\
\!\!&-&\!\!\frac{\dot{\hat{C}}(0)}{2}\phi_{\mu}
\left(\frac{\partial{\cal L}}{\partial\phi}\frac{\partial^{2} {\cal L}}{\partial \phi\partial\phi_\mu}
-\frac{\partial{\cal L}}{\partial\phi_\mu}\frac{\partial^{2} {\cal L}}{\partial \phi^2}\right)\nonumber\\
\!\!&+&\!\!\frac{\dot{\hat{C}}(0)}{2}\phi_{\mu}\phi_{\nu}
\frac{\partial^2{\cal L}}{\partial\phi_\mu \partial\phi}\frac{\partial^{2} {\cal L}}{\partial\phi_\nu \partial\phi} 
\ee
which, when further truncated to the subspace quadratic in $\phi_\mu$, coincides with the
$J_{\lambda\mu,0}=0$ version of \Eqref{eq:flowLquadratic}.

\section{Hamiltonian representation}
\label{sec:Hamiltonian}

The standard way of dealing with flow equations like the spatial integrals of \Eqref{eq:flowL} and \Eqref{eq:flowLsimpler}
is to project them on progressive powers of the Fourier momenta with the help of functional derivatives
in the field, i.e. a DE. We suggest here a different treatment.
If we consider $\phi_M(x)$ to be $x$-independent, we are forced to have all the $\phi_N(x)$, with $N\neq M$,
 $x$-dependent accordingly. In particular, setting
$\phi(x)$ at a constant value forces $\phi_N(x)=0, \ \forall x ,\ \forall N\neq 0$. 
In order to overcome this technical difficulty, we look for a Hamiltonian translation of this equation.
Just as in the traditional Hamiltonian formalism it makes sense to consider $\phi(x)$ and $\pi(x)=\frac{\partial \cal L}{\partial \partial_0\phi(x)}$ to be simultaneously
constant, we want to employ a generalized covariant Hamiltonian formalism that enables us to consider $\phi(x)$ and $\pi^M(x)=\frac{\partial \cal L}{\partial \partial_M\phi(x)}$,
 $M\neq 0$, simultaneously constant. This could be understood as a covariant version of the Ostrogradsky formalism. 
One goes to `phase space' by means of the generalized transform
\be
{\cal H}(x,\phi,\pi^M)={\mathrm ext}_{\phi_M}\Big\{i\pi^M\phi_M+{\cal L}(x,\phi,\phi_M)\Big\}
\ee
and here and in the following the assumed sums over repeated multi-indices do not include the empty index, i.e. $L,M,N\neq0$.
Thus
\begin{align}
\pi^M(x)=i\frac{\partial \cal L}{\partial\phi_M}(x)\quad , \quad \phi_M(x)=-i\frac{\partial \cal H}{\partial \pi^M}(x)\nonumber\\
\frac{\partial \cal H}{\partial \phi}(x)=\frac{\partial \cal L}{\partial \phi}(x)\quad , \quad\frac{\partial \cal H}{\partial x^\mu}(x)=\frac{\partial \cal L}{\partial x^\mu}(x)
\end{align}
and also
\begin{align}
\frac{\partial \pi^L}{\partial\phi_M}(x)
=i\left(\frac{\partial^2 \cal L}{\partial \phi_. \partial \phi_.}\right)^{LM}\!\!\!\!\!\!\!\!\!\!(x)
=i\left(\frac{\partial^2 \cal H}{\partial \pi^. \partial \pi^.}\right)^{-1\, LM}\!\!\!\!\!\!\!\!\!\!\!\!\!\!\!\!(x)\nonumber\\
\frac{\partial^2 \cal L}{\partial \phi_M \partial \phi}(x)=
i\frac{\partial^2 \cal H}{\partial \pi^L \partial \phi}(x)\left(\frac{\partial^2 \cal H}{\partial \pi^. \partial \pi^.}\right)^{-1\, LM}\!\!\!\!\!\!\!\!\!\!\!\!\!\!\!\!(x)\ .
\end{align}
By means of these formulas one can translate the flow equations for ${\cal L}$ into flow equations for ${\cal H}$.

The equations of motion in the Hamiltonian form read
\begin{eqnarray*}
i(-)^M\frac{\mathrm{d}\pi^M}{\mathrm{d}x^M}(x)=\frac{\partial \cal H}{\partial \phi}(x)\\
\phi_M(x)=-i\frac{\partial \cal H}{\partial \pi^M}(x)\ .
\end{eqnarray*}
Notice that the momenta are related to the derivatives of $\phi$ only on-shell, that is on the stationarity trajectories in phase-space,
while off-shell they are independent of them. This enables us to set $\phi$ and all $\pi$'s equal to different and arbitrary constant values.
These constants would become tied to each other if we tried to satisfy the equations of motion by means of constant $\pi^M$ and $\phi$,
in which case we would find the requirement that they correspond to stationarity points of $\cal H$.
The stationarity condition for the momenta, which corresponds to homogeneous $\phi$ configurations,
is expected to be usually solved by setting all the momenta to zero: $\pi^M=0$, purely on the grounds of 
rotational symmetry that enforces contraction of indices.
Nevertheless, this trivial stationarity point might or might not correspond to an absolute minimum.
In case $\cal H$ has an absolute minimum for non-vanishing values of some momenta, one faces a situation where rotational 
symmetry is spontaneously broken~\cite{Lauscher:2000ux} .

Now we want to discuss possible approximations of the flow equation for ${\cal H}$.
If we neglect the explicit spacetime, i.e. momentum, dependence of the couplings,
then ${\cal H}$ depends on the position $x$ only through the fields  $\phi(x)$ and $\pi^M(x)$,
and we can study the flow of the Hamiltonian density by setting both fields to constant values.
Under this approximation, the RG flow of the effective Hamiltonian density is enconded in a
partial differential equation for a function of infinitely many fields. These fields are symmetric tensors
with arbitrarily high rank, and therefore contain higher-spin components.
Here one could consider a further systematic approximation scheme,
that arises by neglecting the dynamics of momenta with rank $M>k$, where $k$ is some chosen positive integer.
That is, in practice, one considers $\cal H$ as independent of these fields.
Other kinds of truncations of the full function $\cal H$ are clearly possible, but will not be discussed in this work.
The zeroth order of such an approximation consists in dropping all the momenta. This is the same as the LPA.
The first order of this expansion originates from keeping only the momentum vector $\pi^\mu$.
For instance \Eqref{eq:flowLsimpler} becomes
\begin{widetext}\vskip-3mm
\be\label{eq:flowHsimpler}
\dot{\cal H}&=&\Lambda^{\eta-2}K_0\left[-\left(\frac{\partial \cal H}{\partial \phi}\right)^2+
\frac{\partial \cal H}{\partial \phi}\frac{\partial \cal H}{\partial \pi^\mu}\frac{\partial^2 \cal H}{\partial \pi^\nu \partial \phi}
\left(\frac{\partial^2 \cal H}{\partial \pi^. \partial \pi^.}\right)^{-1\ \nu\mu}\!\!
+\pi^\mu\frac{\partial \cal H}{\partial \pi^\mu}\frac{\partial^2 \cal H}{\partial \phi^2}
-\left(\frac{\partial \cal H}{\partial \pi^\mu}\frac{\partial^2 \cal H}{\partial \pi^\nu \partial \phi}
\left(\frac{\partial^2 \cal H}{\partial \pi^. \partial \pi^.}\right)^{-1\ \nu\mu}\right)^2
\right]
\nonumber\\
\!\!&&\!\!+\ \Lambda^{d-2+\eta}I_0\frac{\partial^2 \cal H}{\partial \phi^2}
+\frac{\Lambda^{d+\eta}}{d}I_1\delta_{\mu\nu}\left(\frac{\partial^2 \cal H}{\partial \pi^. \partial \pi^.}\right)^{-1\ \nu\mu}
\ee
\end{widetext}
where ${\cal H}$ depends on $\phi$ and $\pi^\mu$ only, i.e. any other momentum with higher rank is neglected.
The Hamiltonian translation of \Eqref{eq:flowLquadratic} instead would miss the last term inside the square bracket of the first line,
and would contain the additional term 
$+\Lambda^{\eta-4}K_1\frac{\partial \cal H}{\partial \pi^\mu}\frac{\partial \cal H}{\partial \pi_\mu}
\left(\frac{\partial^2 \cal H}{\partial \phi^2}\right)^2$.
Because of rotational (or Lorentz) symmetry, an $x$-independent Hamiltonian density in generic space dimensionality $d$ can depend only on two 
scalar variables:  $\phi$ and $\varpi\equiv(\pi^\mu\pi_\mu)/2$. Inserting ${\cal H}(\varpi,\phi)$ into the previous equation and explicitly inverting the $d\times d$ matrix
$\frac{\partial^2 \cal H}{\partial \pi^\mu \partial \pi^\nu}$ one gets
\begin{widetext}
\be\label{eq:flowHfinal}
\dot{\cal H}&=&\Lambda^{\eta-2}K_0\left[-{\cal H}^{(01)\, 2}
+2\varpi{\cal H}^{(10)}\left(\frac{{\cal H}^{(01)}{\cal H}^{(11)}}{{\cal H}^{(10)}+2\varpi{\cal H}^{(20)}}
+{\cal H}^{(02)}\right)
-\left(\frac{2\varpi{\cal H}^{(10)}{\cal H}^{(11)}}{{\cal H}^{(10)}+2\varpi{\cal H}^{(20)}}\right)^2
\right]\nonumber\\
\!\!&&\!\!+\ \Lambda^{d-2+\eta}I_0{\cal H}^{(02)}
+\frac{\Lambda^{d+\eta}}{d}I_1\left(\frac{d-1}{{\cal H}^{(10)}}+\frac{1}{{\cal H}^{(10)}+2\varpi {\cal H}^{(20)}}\right)\ .
\ee
\end{widetext}
Notice that by appropriate rescalings of ${\cal H}, \varpi , \phi ,$ one can absorb (i.e. set equal to one) the three regulator-dependent parameters 
$K_0, I_0, I_1,$ and get a regulator-independent
flow equation. In details, the rescaling is
\be
{\cal H}&\to& a {\cal H}\ ,\quad \quad  \varpi \to b \varpi \  , \quad\ \quad \phi \to c \phi\ ,\nonumber\\
a&=&\frac{I_0}{K_0}\ ,  \quad\quad b=\frac{I_0^2}{K_0^2  I_1}\ ,  \quad \ \  c=\sqrt{I_0}\ . \nonumber
\ee
This is not possible for \Eqref{eq:flowLquadratic} where the last term 
$+\Lambda^{\eta-4}K_1\frac{\partial \cal H}{\partial \pi^\mu}\frac{\partial \cal H}{\partial \pi_\mu}$ 
introduces an extra regulator-dependent
coefficient that, after the previous rescalings, provides an essentially arbitrary parameter $B>0$ multiplying
$2\varpi\left({\cal H}^{(10)}{\cal H}^{(02)}\right)^2$.

\section{3D Ising critical $\eta$}
\label{sec:critical}

We now want to start testing the quality of the truncation encoded in \Eqref{eq:flowHfinal}.
Such an approximation cannot be expected to perform equally well for any kind of observable. In this section we specifically
ask how does this approximation perform in the task of describing the critical properties of a three dimensional theory.
The motivation for starting with this application is that such a critical behavior, namely the possible set of conformal theories at the
phase transitions and the corresponding critical exponents, is well known and provide a traditional benchmark for any nonperturbative
tool in statistical field theory.
To address this question, the first step to take is shifting our attention from the dimensionful fields and interactions to 
the renormalized dimensionless ones. In other words, a conformal behavior of the system is expected to reveal itself by means of self-similarity, 
such that every dimensionful quantity at criticality should scale with $\Lambda$ according to its full (quantum) dimensionality.
We called $d_\phi$ and $d_\pi$ such dimensions for $\phi$ and $\pi^\mu$ respectively; it is therefore convenient to rescale
\be
{\cal H}&\to \Lambda^d {\cal H}\ , \quad  \varpi \to \Lambda^{2 d_\pi} \varpi \  , \quad \phi \to \Lambda^{d_\phi}  \phi
\ee
because then the new quantities can be considered $\Lambda$-independent at criticality.
The full dimensionality of $\pi^\mu$ can be fixed by asking that the Legendre transform term $\pi^\mu\phi_\mu$ have dimension $d$.
This is equivalent to demanding that $(d-d_\phi-1)$ be equal to the full dimension of $\pi_\mu$, as one would expect by performing the 
Legendre transform of a simple Langrangian with a kinetic term of the form $\phi^\mu\phi_\mu/2$. As a consequence we set
$d_\pi=(d-\eta)/2$ and $d_\phi=(d-2+\eta)/2$.

To sum up, the critical theories will be independent of the scale $\Lambda$ and therefore correspond to FPs of the RG flow,
that is, to solutions of the previous equations where one sets $\dot{\cal H}=0$,  $\Lambda=1$ and adds the
canonical rescaling terms:
$d {\cal H}-\left(d-\eta\right)\varpi{\cal H}^{(1,0)}-\left(d-2+\eta\right)\!/2\  \phi{\cal H}^{(0,1)}$, on the r.h.s..
We are interested in studying the simple truncation that arises by projecting the flow equations on the ansatz
\be
{\cal H}(\varpi,\phi)=\varpi/Z(\phi)+V(\phi)
\ee
which corresponds to the Legendre transform of
\be
{\cal L}(\phi_\mu,\phi)=Z(\phi)\frac{\phi_\mu\phi^\mu}{2}+V(\phi)\ .
\ee
Then \Eqref{eq:flowHfinal} provides
\be
\dot{V}\!&=&\! dV-\frac{d-2+\eta}{2}\phi V'-\left(V'\right)^2+V''+Z\label{eq:flowV}\\
\dot{Z}\!&=&\! -\eta Z-\frac{d-2+\eta}{2}\phi Z' -2ZV''+Z''-2\frac{\left(Z'\right)^2}{Z}\label{eq:flowZ}\, .\ \ \ 
\ee
The Hamiltonian translation of \Eqref{eq:flowLquadratic} would lead to
the same equation for $\dot{V}$ and it would add a term $+2B(V''/Z)^2$
on the r.h.s of $\dot{Z}$.
These equations differ from the ones obtained by a first order of the DE~\cite{Ball:1994ji}
in several respects.
The equation for $\dot{V}$ is essentially the same, apart for the fact that in the DE
there is a regulator dependent coefficient, usually identified with the coefficient of $Z$, that cannot
be removed by rescalings. Here instead this can be safely set equal to one.
In \Eqref{eq:flowZ} the first four terms are present also in the DE, even if we 
set the coefficient of the $(ZV'')$-term equal to $-2$ instead of $-4$.
Furthermore the DE includes two terms, $+4V''$ and $+2BV''$, that are absent here.
The latter is the DE-version of the $+2B(V''/Z)^2$ which would be present in the Hamiltonian 
translation of \Eqref{eq:flowLquadratic}.
The last term of \Eqref{eq:flowZ} does not appear in the DE.

We start analyzing these flow equations in the simple approximation of a field-independent $Z$.
This is an admissible solution of the system of FP Eqs.(\ref{eq:flowV},\ref{eq:flowZ})
only for the two trivial cases of the Gau\ss ian FP ($V(\phi)\to 0$, $Z(\phi)^{-1}\to 0$) 
and of the high-temperature FP ($V(\phi)\to (1-\eta/2)\phi^2/2$, $Z(\phi)^{-1}\to 0$) .
In the general case such an approximation can be interpreted as the zeroth order 
of a polynomial truncation of $Z$ around some point $\phi_*$.
The corresponding FP equation for $Z(\phi_*)$ cannot be solved for the latter,
since it entails
\be
\eta=-2V''(\phi_*)\ .
\ee
As far as the choice of $\phi_*$ is concerned, it is to be expected that the best polynomial approximations
come from the choice of stationarity points for the potential $V$.
For the Wilson-Fisher FP we know in advance that there are two such points, $\phi_*=0$ with 
$V''(\phi_*)<0$ and $\phi_*\neq0$ with $V''(\phi_*)>0$.
The latter would give $\eta<0$ and is therefore a bad choice in the present case.
The former leads to $\eta=0.19853$, which can be obtained by the numerical technique 
described later on in this Section, or by shooting from the origin with $V'(0)=0$ and $V''(0)=-\eta/2$.

The approximation of constant $Z$ is clearly too poor to provide a fair estimate of $\eta$,
yet it enjoys the good property that $Z(0)$ remains a free parameter, which corresponds
to the wanted invariance under rescaling of the field.
Whether this remains true in a less severe approximation is a question of practical importance,
since it affects the soundness and uniqueness of the estimate of the critical exponents.
If we move to higher polynomial orders in $Z(\phi)$, the explicit independence from $Z(0)$
is lost. For instance, upon inclusion of $Z(\phi)^{-1}=\zeta_0(1+\zeta_1\phi^2/2)$
one finds
\be
\eta=-2V''(0)\frac{2+\zeta_0}{2-\zeta_0}
\ee
and a similar parametric dependence survives also including one more coupling in $Z$.
Since it does not seem straightforward to go to higher orders of 
this kind of successive approximations,
let us turn to the task of solving the full system of Eqs.(\ref{eq:flowV},\ref{eq:flowZ}) at 
a FP.

We choose the method of shooting from large field values~\cite{Morris:1997xj}.
To this end we use the large-field asymptotics of the FP solution, as parametrized by
$\eta$ itself and other two free parameters $A_V$ and $A_Z$.
The reader can find it in Appendix~\ref{sec:appendix}.
These three parameters are not completely free since 
the FP solutions we are after enjoy
$\mathbb{Z}_2$-symmetry, that is $V'(0)=Z'(0)=0$.
Thus, we need numerical solutions that interpolate the
right field asymptotics, say at $\phi=1$, with the needed behavior at $\phi=0$.
Unfortunately, regardless the use of high-order large field asymptotic expansions (ten terms for each function),
the numerical integration does not always reach $\phi=0$.
In general, one gets a solution that extends till $\phi=0$ only if the corresponding
$V'(0)$ and $Z'(0)$ would be small enough.
Then, one could trade in the parameter $A_Z$ for $Z(0)$, even if this is not necessary.
At fixed generic values of $\eta$ and $A_V$, $Z(0)$ is bigger if $A_Z$ is closer to zero.
The two parameters have the same sign and roughly their order of magnitude is related 
by $Z(0)\sim (10A_Z)^{-1}$.

\begin{figure}[!t] 
\begin{center}
\includegraphics[width=0.2385\textwidth]{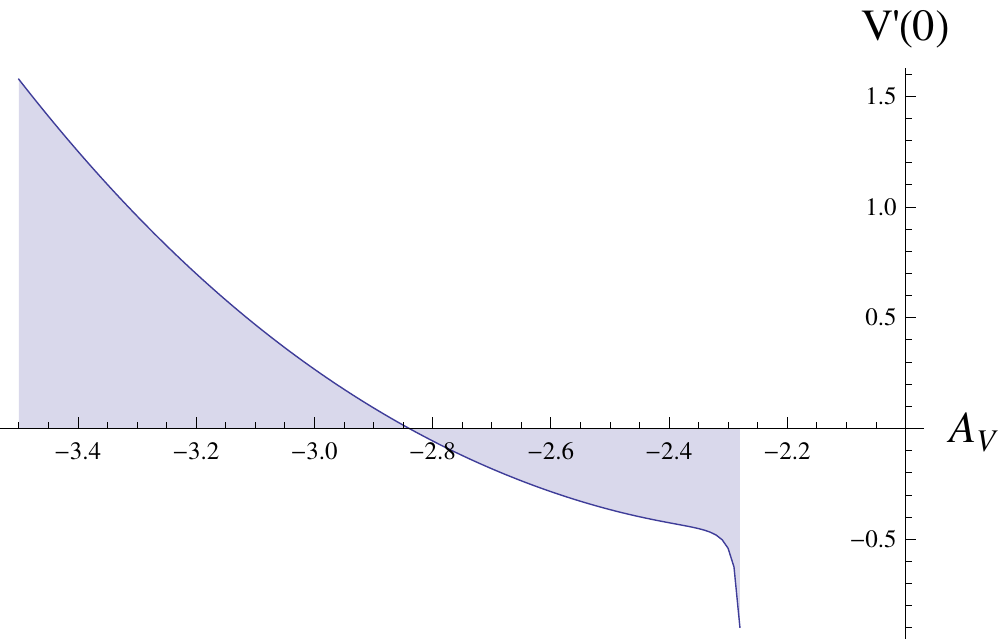}
\includegraphics[width=0.2385\textwidth]{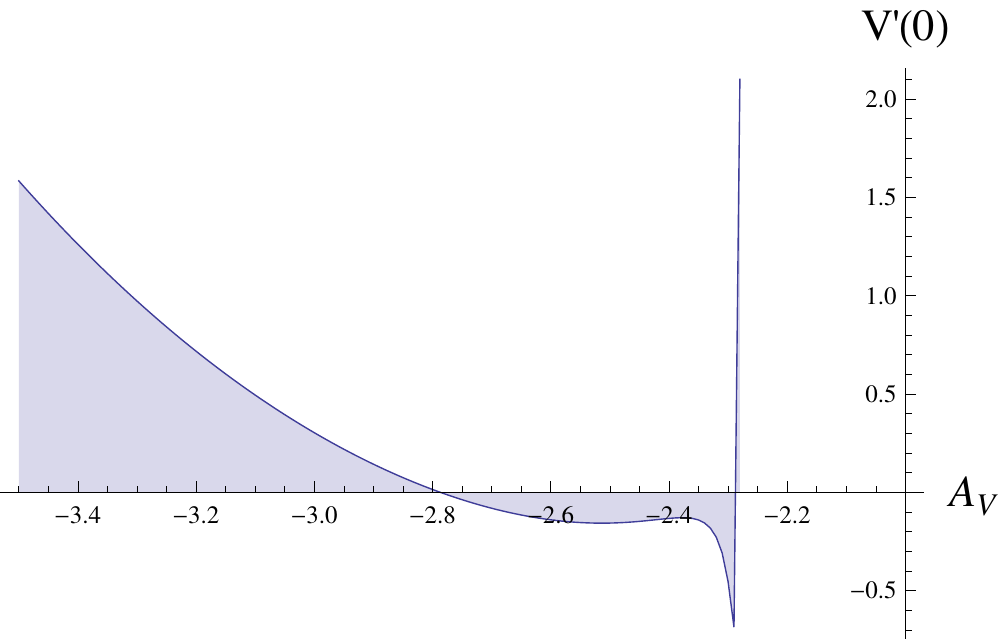}\quad
\includegraphics[width=0.2385\textwidth]{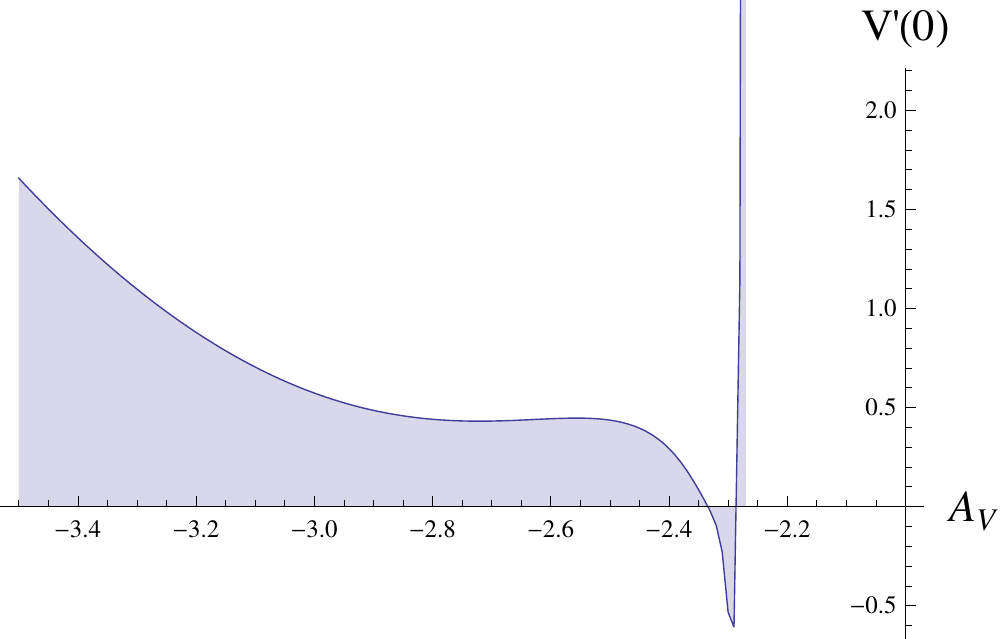}
\includegraphics[width=0.2385\textwidth]{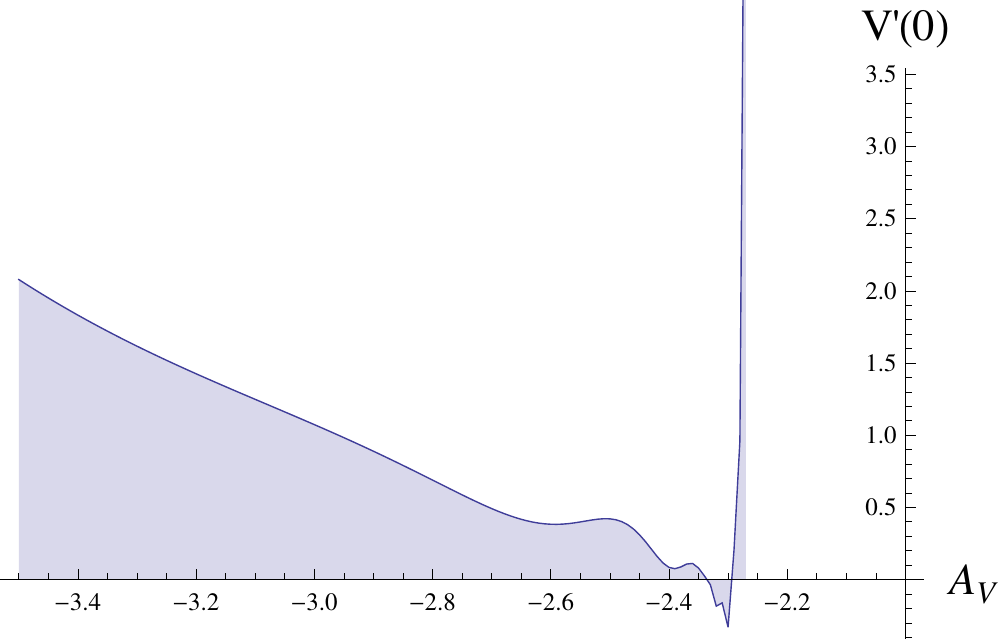}\quad
\includegraphics[width=0.2385\textwidth]{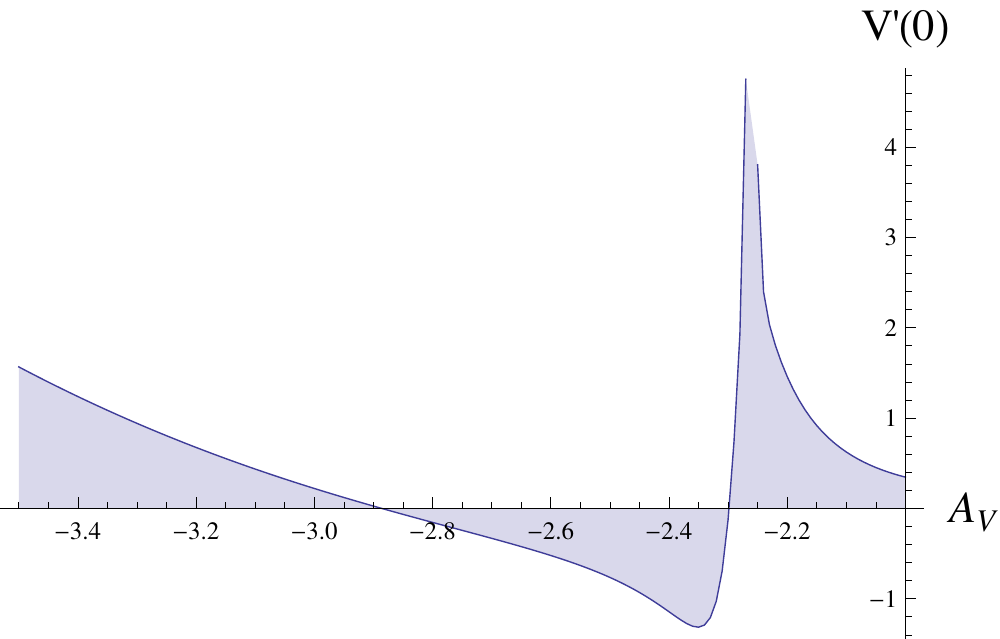}
\includegraphics[width=0.2385\textwidth]{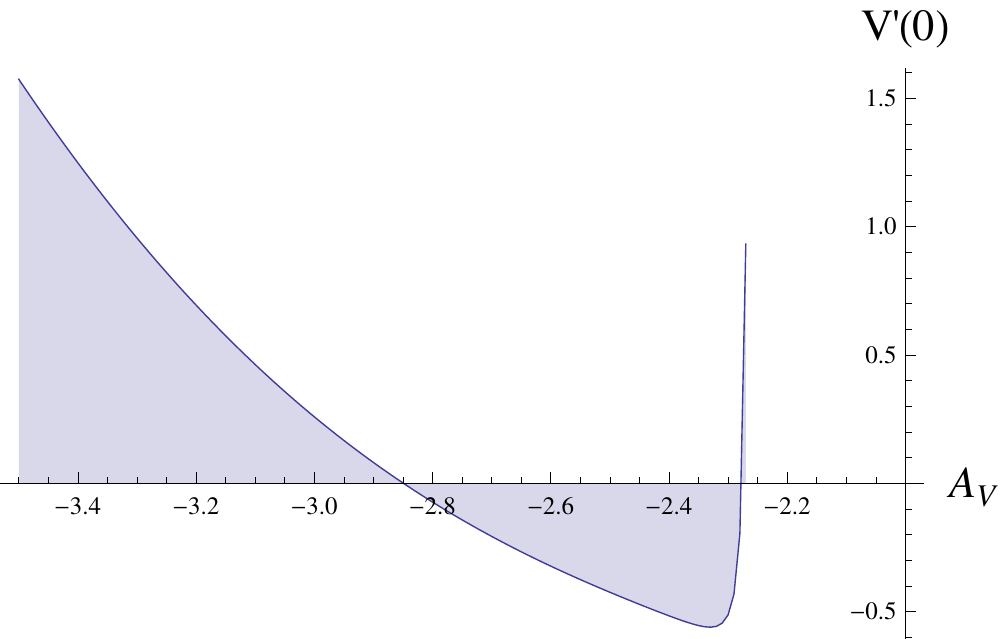}
 \caption{$V'(0)$ as a function of $A_V$, at fixed $\eta=1.2\times 10^{-2}$ and for
various values of $A_Z\in\{1, 10^{-1}, 10^{-2}, 10^{-3}, -10^{-1}, -1\}$
(from left to right in each row, and from top row to bottom row).
These are discrete plots with step-size $\Delta A_V=10^{-2}$. 
Where the curve is missing, the numerical integration 
(starting from $\phi=1$) does not reach $\phi=0$. 
}
\label{fig:dV0_eta0p012}
\end{center}
\end{figure}
\begin{figure}[!t] 
\begin{center}
\includegraphics[width=0.2385\textwidth]{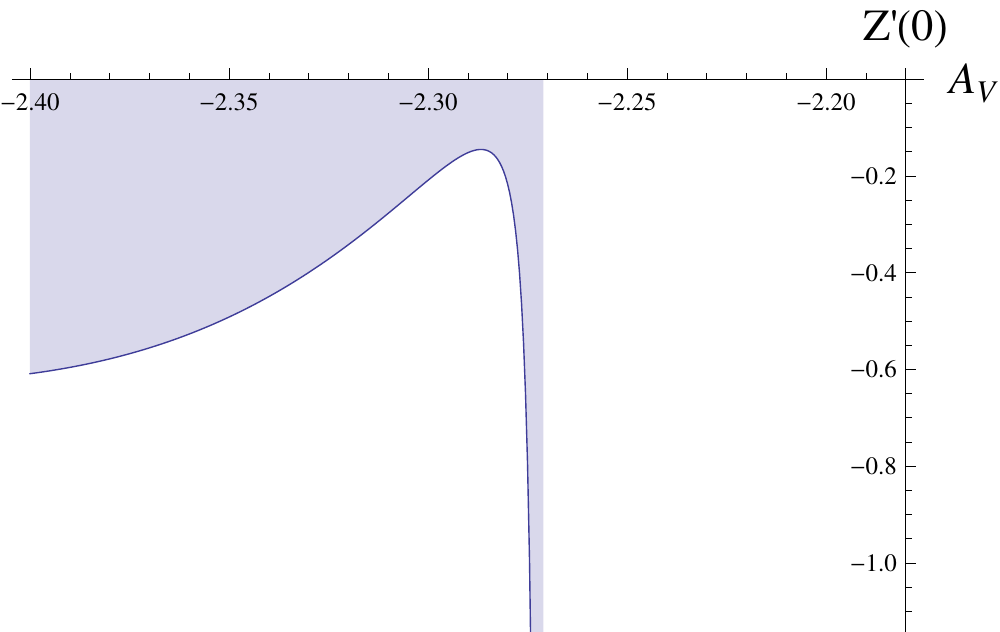}
\includegraphics[width=0.2385\textwidth]{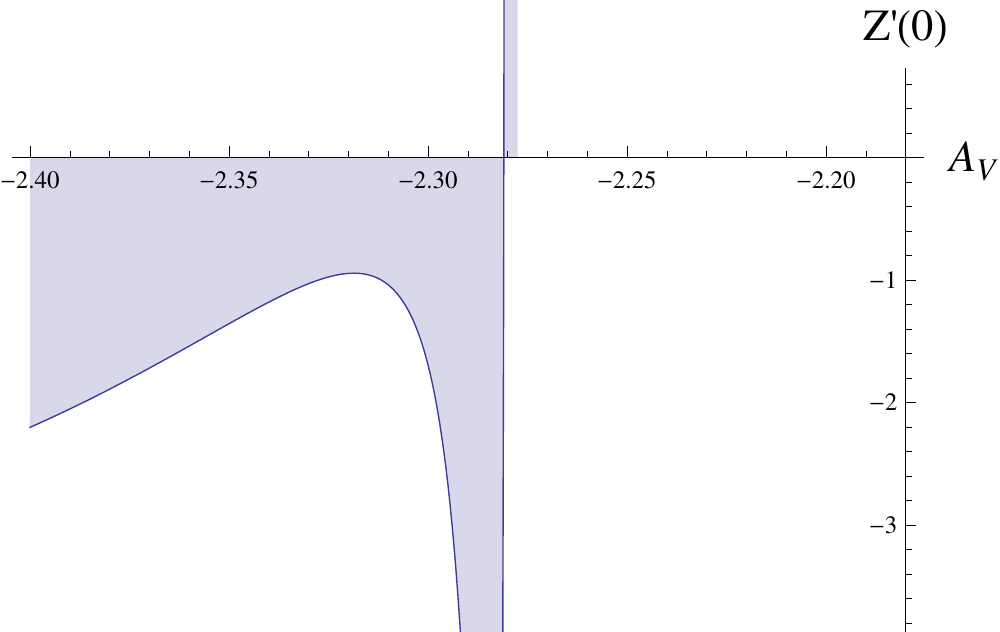}\quad
\includegraphics[width=0.2385\textwidth]{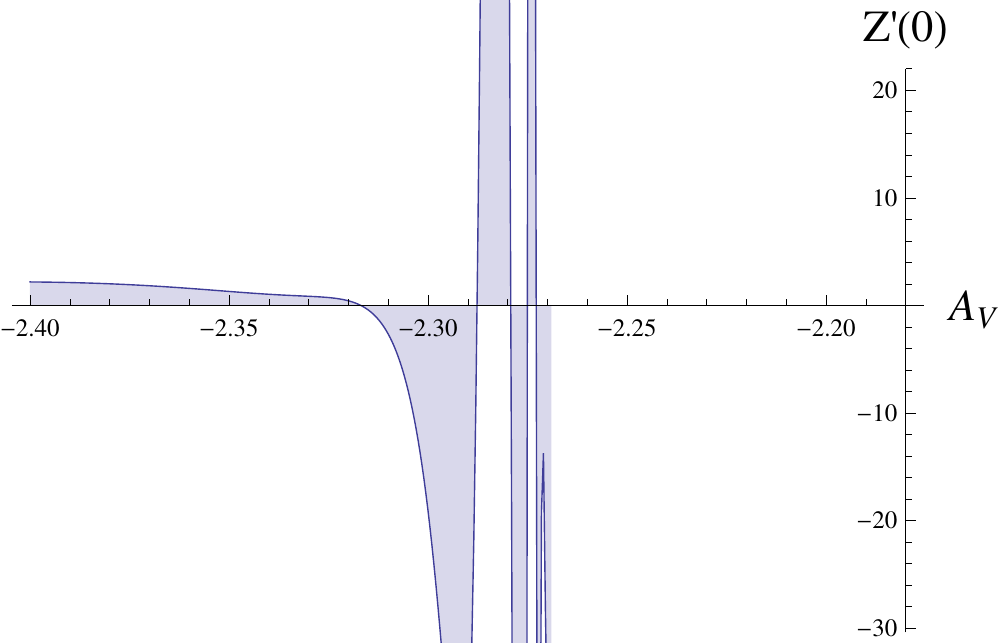}
\includegraphics[width=0.2385\textwidth]{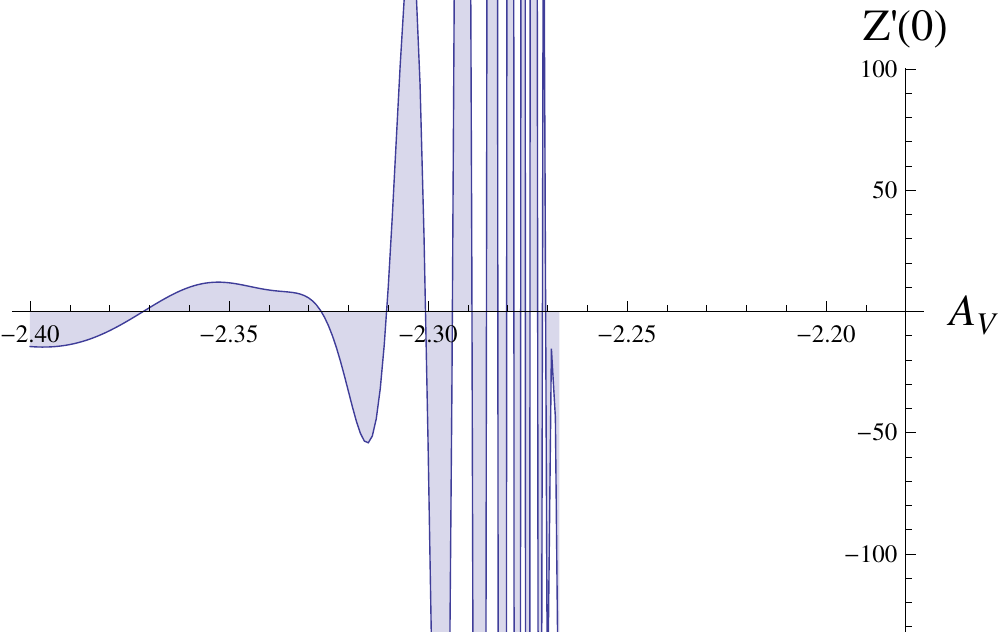}\quad
\includegraphics[width=0.2385\textwidth]{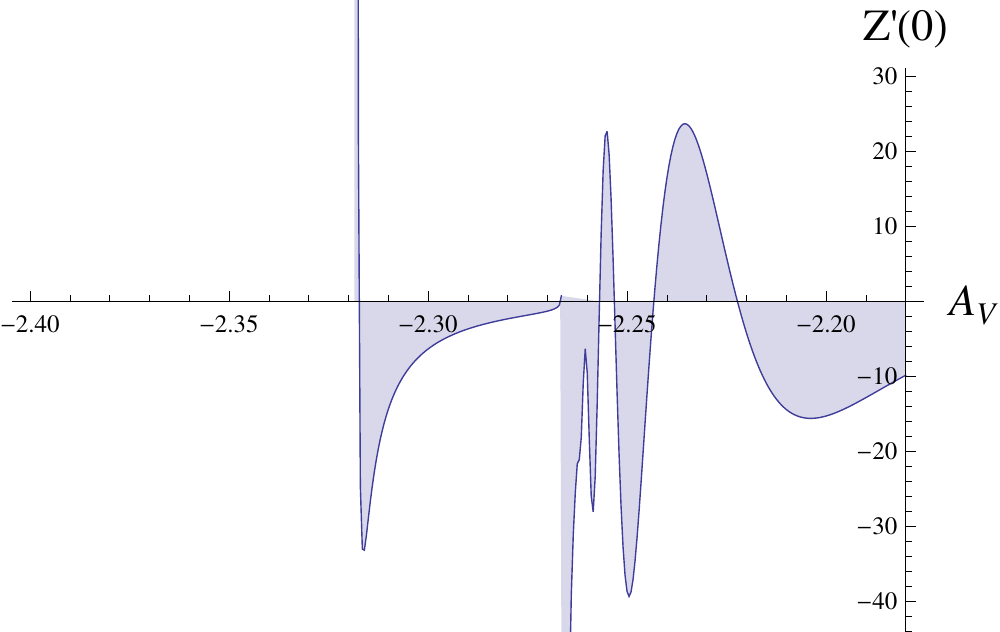}
\includegraphics[width=0.2385\textwidth]{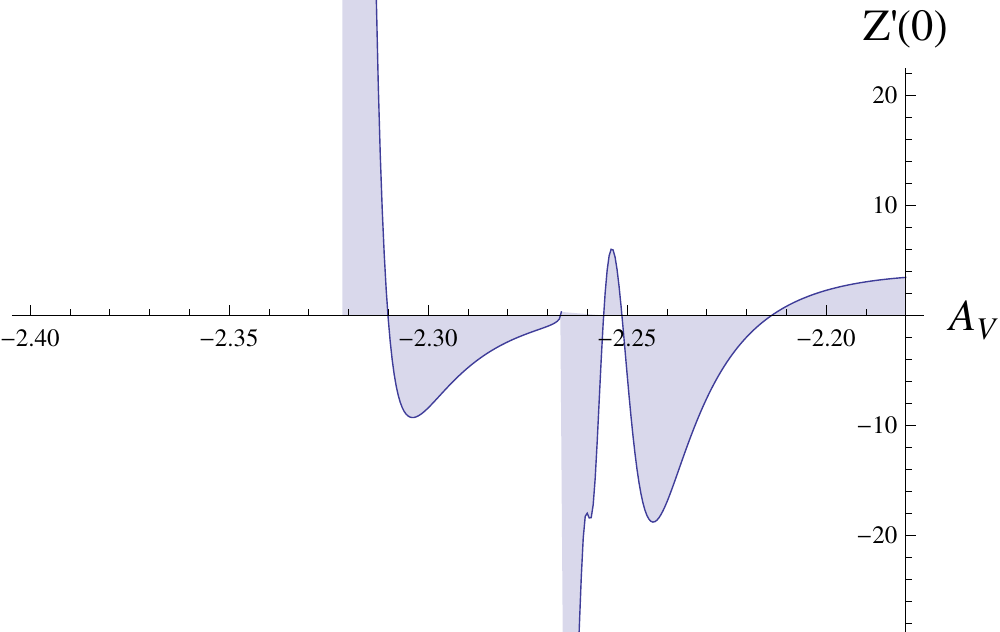}
\includegraphics[width=0.2385\textwidth]{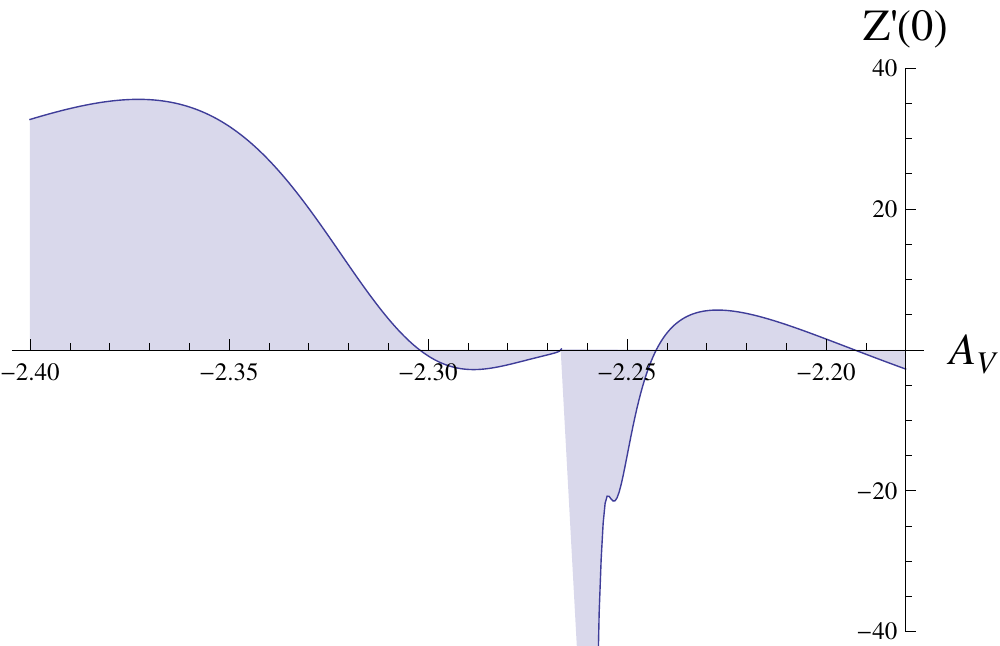}
\includegraphics[width=0.2385\textwidth]{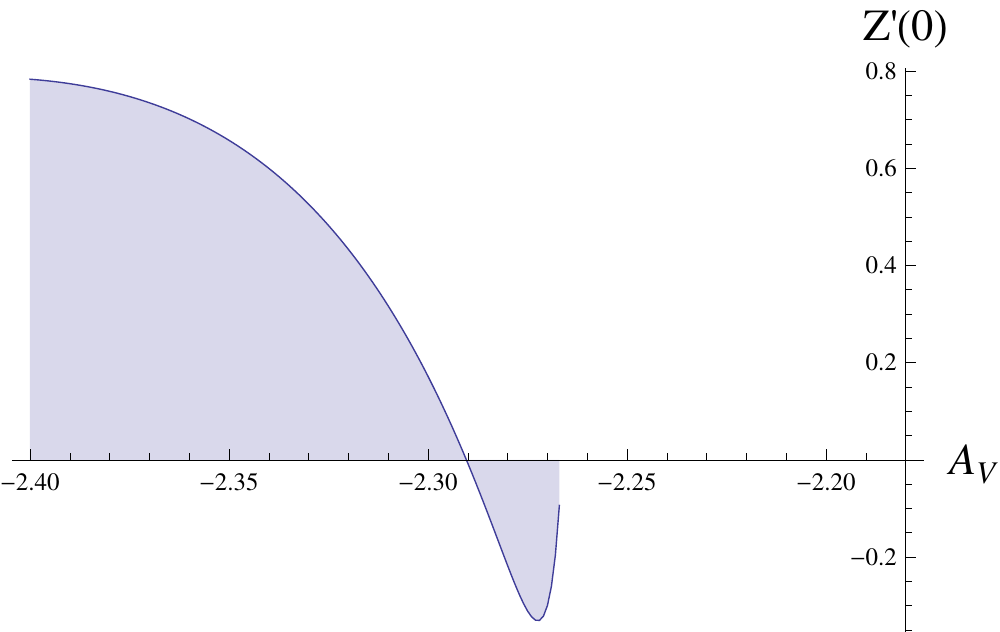}
 \caption{$Z'(0)$ as a function of $A_V$, at fixed $\eta=1.2\times 10^{-2}$ and for
various values of 
$A_Z\in\{1, 10^{-1}, 10^{-2}, 10^{-3}, -10^{-2}, -3\times 10^{-2}, -10^{-1},  -1\}$
(from left to right in each row, and from top row to bottom row).
These are discrete plots with step-size $\Delta A_V=5\times 10^{-4}$. 
Where the curve is missing the numerical integration 
(starting from $\phi=1$) does not reach $\phi=0$. 
}
\label{fig:dZ0_eta0p012}
\end{center}
\end{figure}

The strategy we are going to follow for constructing the FP solutions is the following.
The symmetry conditions at the origin can be used to fix $A_V$ and $\eta$, 
namely by locating a discrete set of points in the $(A_V,\eta)$-plane where both $V'(0)$ and $Z'(0)$ vanish.
In so doing $A_Z$ remains undetermined.
Thus, by variation of $A_Z$ and relocation of the zeros in $(A_V,\eta)$ one can construct  a discrete set of lines of FPs.
For practical reasons, it is mandatory to 
analyze the $A_Z$-dependence of $V'(0)$ and $Z'(0)$ even before extracting
their common zeros in the $(A_V,\eta)$-plane.
This is because the number of their zeros changes as $A_Z$ is changed.

Then we start by fixing an arbitrary initial $\eta$, for instance $\eta=1.2\times 10^{-2}$,
 and plot $V'(0)$ and $Z'(0)$ as functions of $A_V$, for several values of $A_Z$.
Because of the numerical shooting procedure, the resolution of these plots is limited,
a fact that represents the main source of uncertainties in the final estimate of $\eta$.
We first analyze $V'(0)$, which is shown in Fig.~\ref{fig:dV0_eta0p012}. 
The numerical integration is successful
in reaching the origin when $A_V$ is close to $-2.5$, where $V'(0)$ shows two zeros.
The left one corresponds to a quadratic potential (far enough from the matching with 
the asymptotic expansion), therefore we expect it to be connected to
the high-temperature FP, while at the right one $V(\phi)$
has the right qualitative shape for a Wilson-Fisher FP.
Unfortunately, the latter is much harder to locate due to the fact that the curve is very steep
close to this zero, such that a high-resolution plot is needed to reveal it.
For instance, in the first (upper left) panel of Fig.~\ref{fig:dV0_eta0p012}
where $A_Z=1$ the resolution of the plot is too low to show this zero.
If we increase $A_Z$ the slope further increases, making this practical problem more severe.
One the other hand, decreasing $A_Z$, i.e. moving to the following panels of 
Fig.~\ref{fig:dV0_eta0p012} , makes the curve less steep and the location of the rightmost
zero easier. However, the number of the zeros and their qualitative position does not change.
We can even lower $A_Z$ to negative values, as in the last two (lower) panels
of Fig.~\ref{fig:dV0_eta0p012} , and the very same two zeros remain visible.
The fact that the zeros of $V'(0)$ are two is however not generic.
For different values of $\eta$ we observe more than these two zeros,
but we interpret the fact that these additional zeros are not present for all values of $A_Z$ 
as a manifestation of their spurious nature.

Still at $\eta=1.2\times 10^{-2}$, the picture for $Z'(0)$ as a function of $A_V$  is more complicated
than the one for $V'(0)$, and it is shown in Fig.~\ref{fig:dZ0_eta0p012}.
At $A_Z=1$, in the upper left panel, there seems to be no zero.
As we move to $A_Z=10^{-1}$, in the upper right panel, a zero becomes visible.
Lowering further $A_Z$ more zeros show up, as in the third and fourth panel of
Fig.~\ref{fig:dZ0_eta0p012} (mid row), revealing that $Z'(0)$ is wildly oscillating
close to the value of $A_V$ beyond which the integration is no longer reaching $\phi=0$.
This observation, together with the previous study of $V'(0)$, 
suggests that even at $A_Z=1$ there can be zeros which are hard to reveal because 
$Z'(0)$ is too steep in their neighborhood.
Thus, expecting that the slope of the curve be again a growing function of $A_Z$,
we are lead towards lowering the latter parameter, even below zero.
For a negative $A_Z$ close enough to zero, like $A_Z=10^{-1}$ in the lower left panel,
there still are oscillations in an inner-region of $A_V$, let us say roughly on the right of the point that for positive $A_Z$ was the end of successful numerical integration.
On the left of such a point, instead, there are only two zeros, which seem to remain isolated and clearly distinguishable independently of $A_Z$.
If $A_Z$ is further lowered, as in the last (lower right) panel where $A_Z=-1$,
the inner oscillations disappear, the curve becomes less steep, and only the latter two
zeros survive.
On the basis of all these facts, we assume that these two zeros exist for any $A_Z$, 
such that when $A_Z$ is increased from $-1$ towards $+1$ they must get closer and closer, while the function $Z'(0)$ itself becomes steeper and steeper in their vicinity.
By inspecting the shape of the corresponding solutions we can again discard one of these
two zeros, the one on the right. In fact, the corresponding $V'(\phi)$ is positive
all over its domain, and again linear far enough from the 
the asymptotic large field behavior. At the left zero instead, $V(\phi)$ has the right shape for a Wilson-Fisher FP.
\begin{figure}[!t] 
\begin{center}
\includegraphics[width=0.45\textwidth]{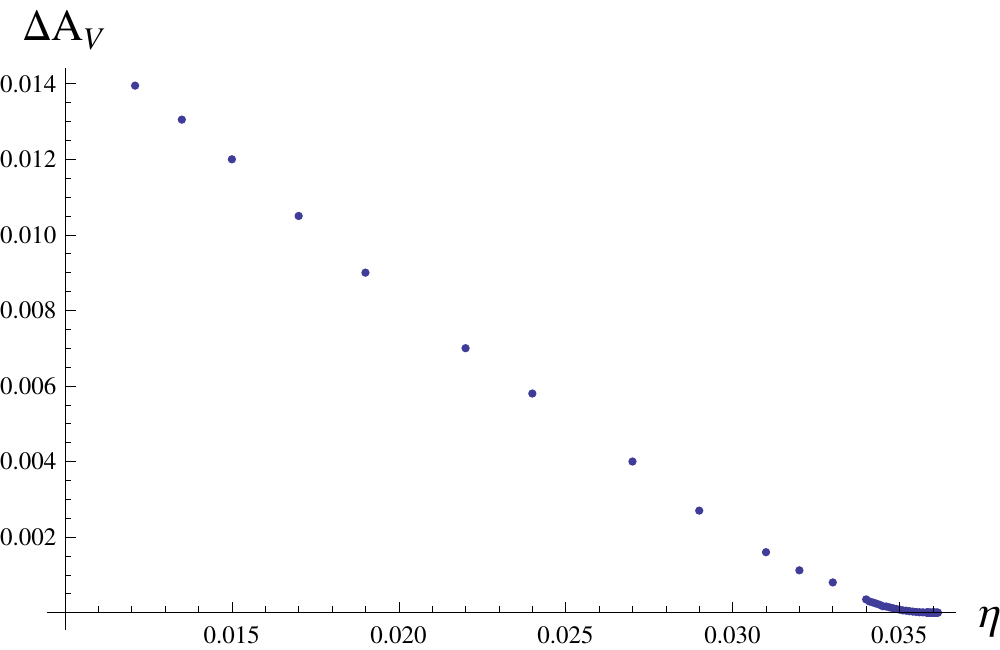}\vskip2mm
\includegraphics[width=0.47\textwidth]{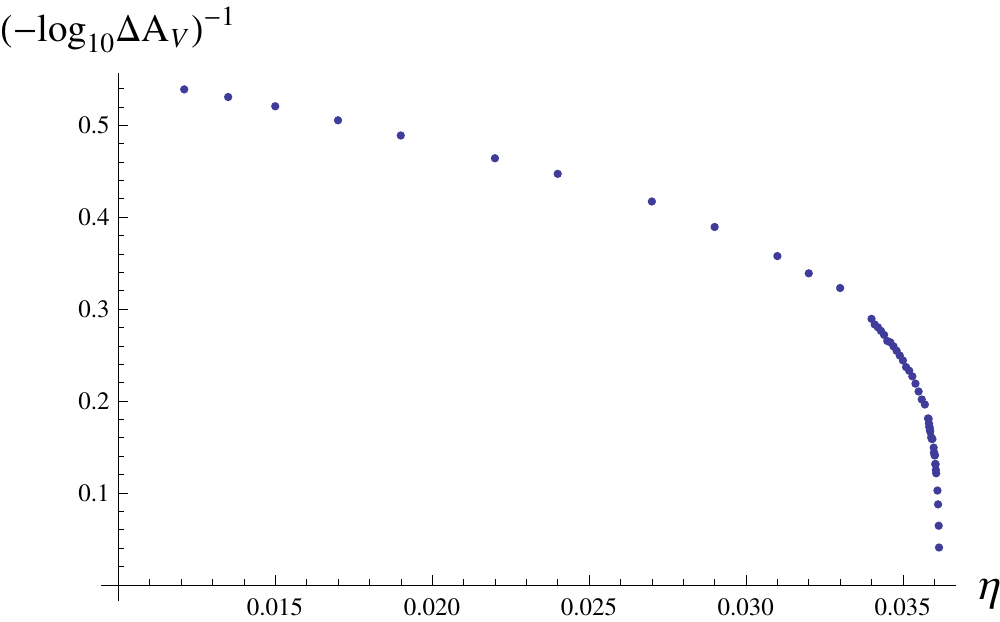}
 \caption{$\Delta A_V$ is the difference between the location of the zero of $Z'(0)$ and that of $V'(0)$,
and here it is considered as a function of $\eta$, at fixed $A_Z=-1$.
}
\label{fig:deltaA_of_eta}
\end{center}
\end{figure}
\begin{figure}[!t] 
\begin{center}
\includegraphics[width=0.45\textwidth]{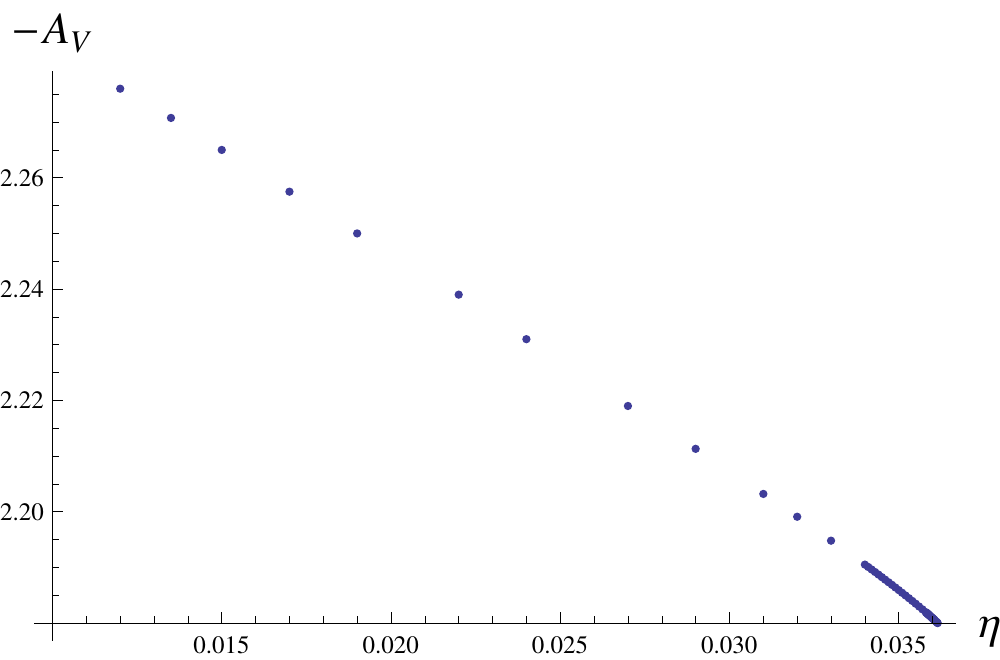}\vskip2mm
\includegraphics[width=0.48\textwidth]{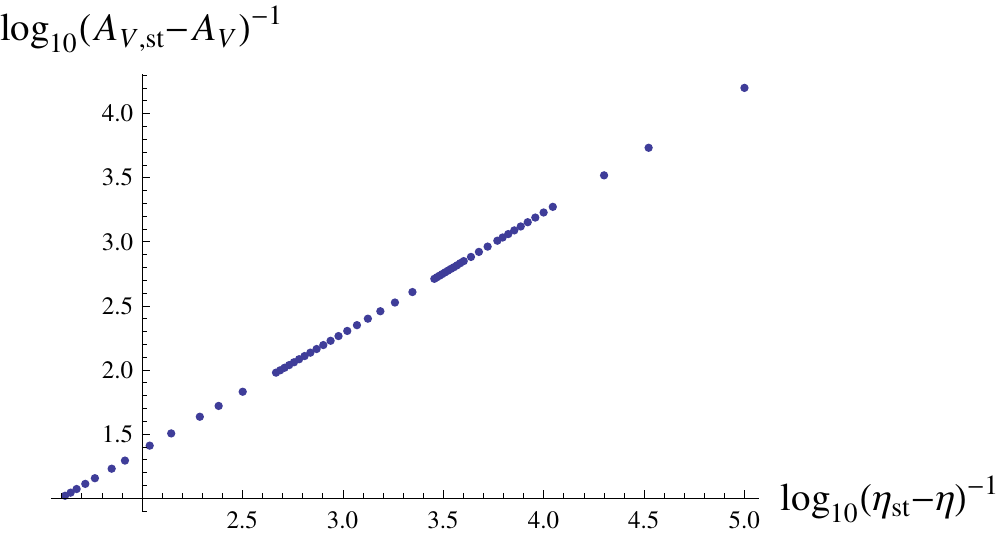}
 \caption{The position of the zero of $V'(0)$ on the $A_V$ axis,
as a function of $\eta$ at fixed $A_Z=-1$, 
in linear (upper panel) and double-logarithmic (lower panel) scale.
In the latter case $\eta_{\text st}$ and $A_{V,{\text st}}$ are the values 
corresponding to the last point on the lists, i.e. the closest to the physical fixed point.
}
\label{fig:A_of_eta}
\end{center}
\end{figure}

To sum up, at the starting value $\eta=1.2\times 10^{-2}$, 
by requiring existence for any $A_Z$ and a proper shape for
the potential $V(\phi)$, compatible with the properties
of the Wilson-Fisher FP, it is possible to
select one zero for $V'(0)$ and one for $Z'(0)$.
Then, the construction of the true FP proceeds by merging these 
zeros by tuning $\eta$.
Having observed that locating the above mentioned zeros is 
easier for lower $A_Z$, especially for negative values, we 
do this at $A_Z=-1$.
When $\eta$  increases, both zeros move towards less negative values.
However, they do so at a different speed, such that they get closer and closer.
Since the zero of $V'(0)$ is always on the right of that of $Z'(0)$,
let us define $\Delta A_V$ as the position of the former minus the one of the latter.
This quantity approaches zero with exponential rate in $\eta$, 
as can be guessed by inspecting the upper panel of Fig.~\ref{fig:deltaA_of_eta}.
We can model this by looking at the function   
$F(\eta)=-(\log_{10}\Delta A_V)^{-1}$, that is   $\Delta A_V=10^{-1/F(\eta)}$,
such that if  $F(\eta_*)=0$ then  $\Delta A_V$ vanishes exponentially
when $\eta\to\eta_*$.
This is indeed the case, as the lower panel of Fig.~\ref{fig:deltaA_of_eta}
reveals.

The last point in this list, at which we stopped the merging procedure,
is $\eta_{\text st}=0.03615$.
At this value of $\eta$ the position of the zero of $V'(0)$ is
$A_{V,{\text st}}=-2.17999226178876724872935061$ 
while there is still a difference of approximately $\Delta A_{V,\text st}=2.4\times 10^{-25}$.
For completeness in  Fig.~\ref{fig:A_of_eta} we also show how the position of
the zero of $V'(0)$ depends on $\eta$.
From this set of data we need to estimate $\eta_*$ by extrapolation.
We assume that the overall shape of the curve sampled in the right panel of Fig.~\ref{fig:deltaA_of_eta}
does not change beyond $\eta_{\text st}$.
This entails that the true zero $\eta_*$ must be bigger than $\eta_{\text st}$,
as well as smaller than the zero obtained by linear extrapolation, which is $\eta_{\text ex}=0.036167$.
We therefore take the mid point in between these two extrema as our reference expectation value, and half of their difference as 
an estimate of the numerical uncertainty. To sum up, at $A_Z=-1$ we find
$\eta_*=(3.616\pm0.001)\times 10^{-2}$.

\begin{figure}[!t] 
\begin{center}
\includegraphics[width=0.45\textwidth]{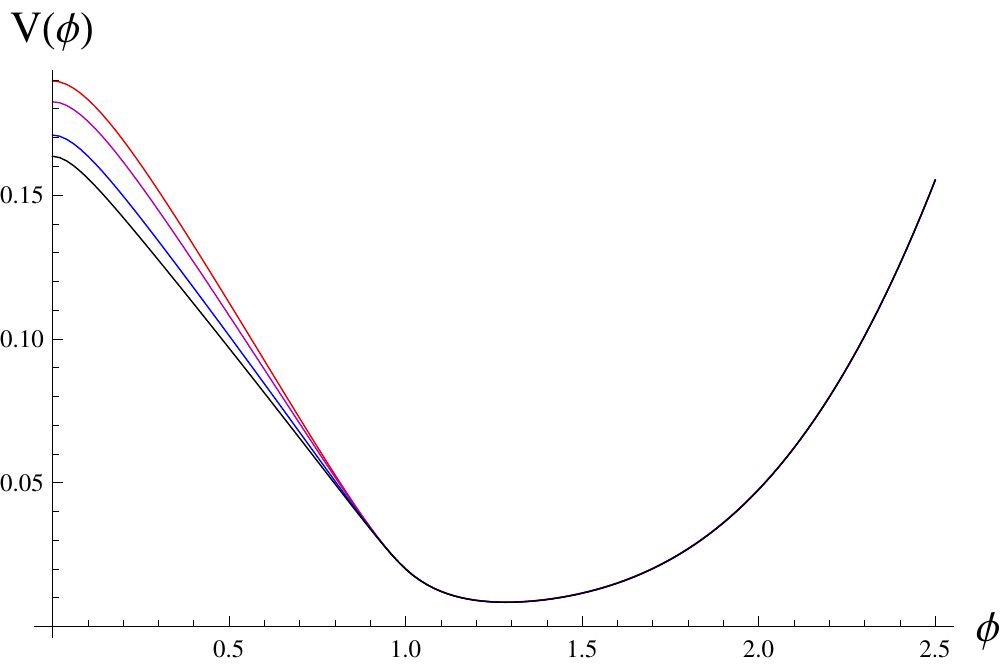}
\includegraphics[width=0.45\textwidth]{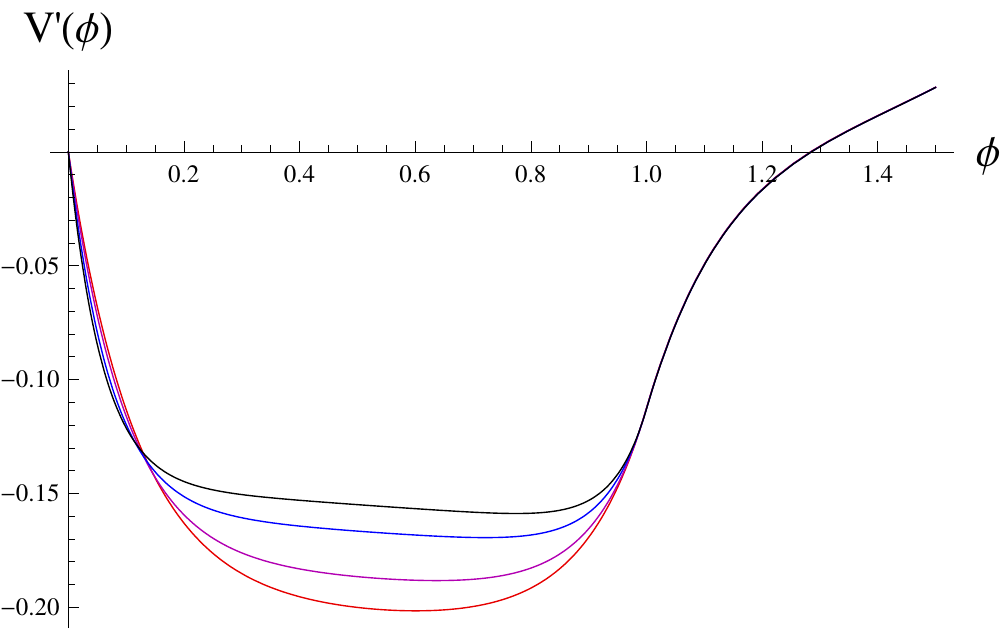}
 \caption{The potential $V(\phi)$ (upper panel) and its first derivative $V'(\phi)$
(lower panel) at the values of $A_V$ corresponding to the zeros of $V'(0)$,
for $A_Z=-1$ and for $\eta\in\{0.03600, 0.03605, 0.03610, 0.03612\}$
from red (deeper) to black (shallower).
All these plots are obtained by numerical integration from $\phi=1$
to $\phi=0$ and are extended beyond $\phi=1$ by means of the
large field asymptotic behavior $V_{\text as}$.
}
\label{fig:V_of_phi}
\end{center}
\end{figure}

\begin{figure}[!t] 
\begin{center}
\includegraphics[width=0.45\textwidth]{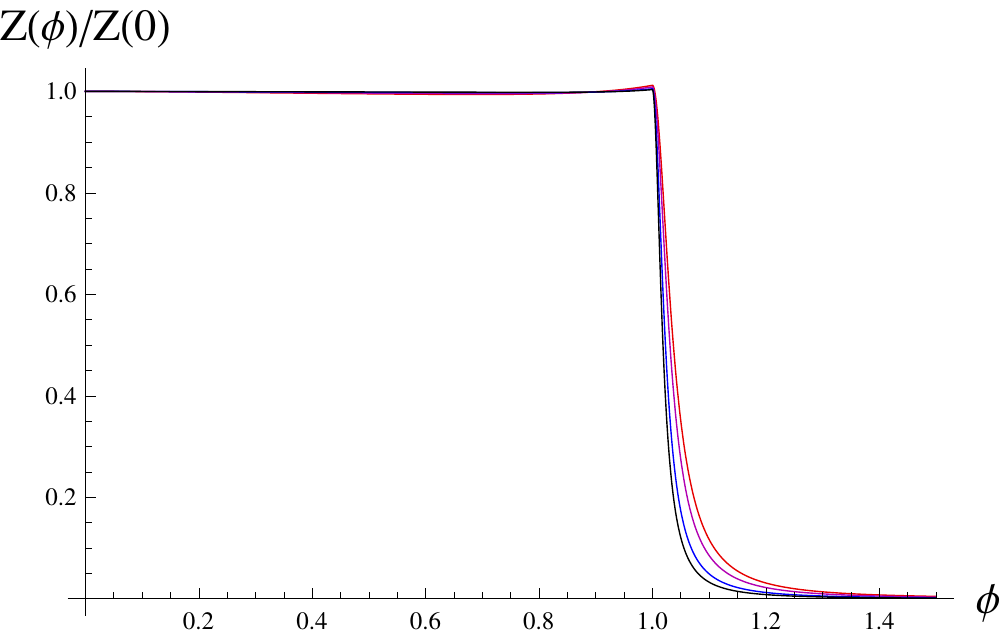}\vskip2mm
\includegraphics[width=0.2385\textwidth]{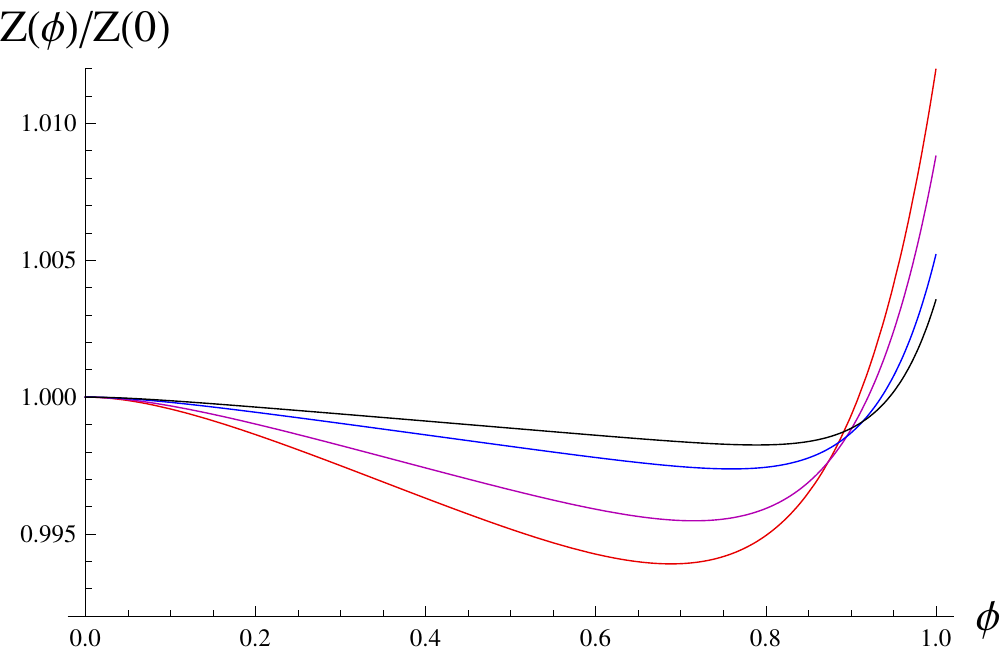}
\includegraphics[width=0.2385\textwidth]{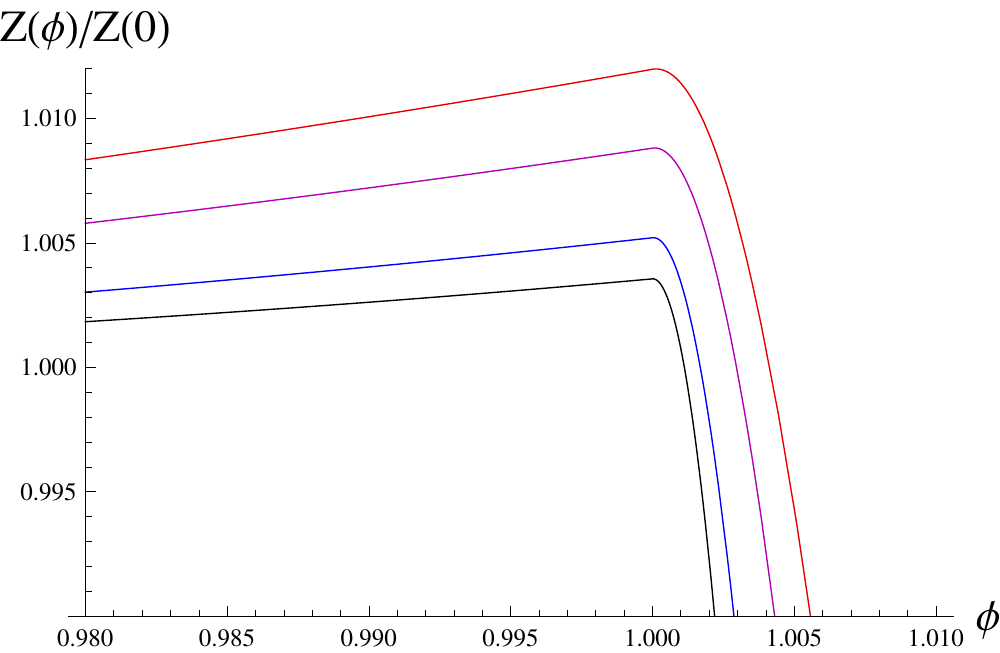}
 \caption{The function $Z(\phi)$, normalized to one at the origin,
at the values of $A_V$ corresponding to the zeros of $Z'(0)$,
for $A_Z=-1$ and for $\eta\in\{0.03600, 0.03605, 0.03610, 0.03612\}$
from red (smoother) to black (sharper).
The true overall scales are 
$Z(0)\in\{-37.6, -53.4, -96.3, -146.8\}$
correspondingly.
All these plots are obtained by numerical integration from $\phi=1$
to $\phi=0$ and are extended beyond $\phi=1$ by means of the
large field asymptotics $Z_{\text as}$.
In the lower panels it is shown the behavior in the inner region (left)
and close to the matching with the outer region (right).
}
\label{fig:Z_of_phi}
\end{center}
\end{figure}

Though the difference between the location of the two zeros is tiny if quantified 
in terms of $\Delta A_V$, it is less satisfactory if quantified in terms of the actual values
of $V'(0)$ and $Z'(0)$ at a random point in between these two zeros.
Take for instance $\eta=\eta_{\text st}$ and inspect the solution at the zero of $V'(0)$:
the $\mathbb{Z}_2$ symmetry of $Z$ is severely violated by $Z'(0)=-234$.
Vice versa, at the zero of $Z'(0)$ one has $V'(0)=-0.114$.
Thus, in order to construct a good numerical approximation of the FP functions
it would be necessary to decrease $\Delta A_V$ much beyond the point where we stopped
our analysis.
Yet, we can get a qualitative portrait of these functions by observing how they evolve along
the curves of the corresponding zeros parameterized by $\eta$.
That is, in Fig.~\ref{fig:V_of_phi} we show the scalar potential $V(\phi)$ (left panel) 
and its derivative (right panel) at the zeros of $V'(0)$ for several values of $\eta$.
For the same set of $\eta$'s the reader can find in Fig.~\ref{fig:Z_of_phi}
the function $Z(\phi)$, normalized by the corresponding $Z(0)$, at the zeros of $Z'(0)$.

Eventually one needs to consider again the effect of changing $A_Z$.
On the basis of our previous discussion, we expect that the
zeros of $V'(0)$ and $Z'(0)$ that we have analyzed at $A_Z=-1$
continue to exist also at any other value of $A_Z$,
while their position should smoothly depend on the latter parameter.
Indeed, it can be immediately observed, for instance by comparing 
the plots in the last line of Fig.~\ref{fig:dV0_eta0p012} to those in
the last line of Fig.~\ref{fig:dZ0_eta0p012}, 
that at $\eta=0.012$ the distance $\Delta A_V$ does indeed change
when we shift $A_Z$ from $-1$ to $-10^{-1}$.
Therefore the curves in Fig.~\ref{fig:deltaA_of_eta} must change.
Yet, this does not prove that the position where these curves meet the
horizontal axis is shifting.
To assess whether this is the case or not, one should produce 
new curves at several other values of $A_Z$, and extrapolate the
position of the critical $\eta$.
This analysis has not been performed so far.
What we tried instead is a simpler check, that can only give us
a bound on the $A_Z$-dependence of the critical $\eta$.
This is related to the linear extrapolation method that we adopted
for the estimate of $\eta$ at $A_Z=-1$.
We focus on two values of $\eta$ that for $A_Z=-1$ are close to the
critical one, such that we can use the corresponding $\Delta A_V$
to linearly extrapolate a best value and a error bar for $\eta$.
We then analyze the $A_Z$ dependence of $\Delta A_V$
at these two points only. If there is any, by extrapolation 
we can estimate the change in the critical $\eta$.
In order to be more precise in revealing possible changes,
we need to be as close as possible to the critical point.
Therefore we chose these two points to be 
$\eta_{\text st}=0.03615$ and $\eta=0.03614$.
Since the changes are expected to be more pronounced
far from the critical point, especially the value of $\Delta A_V$
at the latter $\eta$ has been under our focus.
For $A_Z=-1$ this is approximately $\Delta A_V=3\times 10^{-15}$.
We then followed the position of the zeros of $V'(0)$ and $Z'(0)$
changing $A_Z$ in small discrete steps. By gradually increasing the 
step-size we  sampled  the interval $A_Z\in[-1,+10^{-3}]$.
We never witnessed any change in $\Delta A_V$.
This leads us to the conclusion that, if the critical $\eta$ undergoes a variation
when $A_Z$ is changed inside  $[-1,+10^{-3}]$,
this variation must be smaller than the numerical uncertainty
of our estimate at $A_Z=-1$.

Unfortunately this analysis has been affected by
all the problematic features 
of the pattern of zeros that we outlined while commenting 
Figs.~\ref{fig:dV0_eta0p012} and \ref{fig:dZ0_eta0p012} .
Specifically, we again observed that the shapes of $V'(0)$ and $Z'(0)$
as functions of $A_V$ become steeper and steeper close to their
zeros, as $A_Z$ increases from $-1$ towards $0$ and then to positive values.
This makes the numerical location of these zeros harder and harder,
such that from some values of $A_Z$ on (around $A_Z=-0.5$) we simply identified
the position of the zeros with the location of the end of successful numerical
integration, as in the upper-left panels of Figs.~\ref{fig:dV0_eta0p012}
and \ref{fig:dZ0_eta0p012} .
A further unpleasant ambiguity affects the study of positive $A_Z$ values,
since in this case there appear wild oscillations of these
functions, similar to those that were depicted  in Figs.~\ref{fig:dV0_eta0p012}
and \ref{fig:dZ0_eta0p012} , which often bring corresponding 
additional zeros. We then stuck to our assumption that the zeros
connected to those at $A_Z=-1$ occur indefinitely close to
the ending point of successful numerical integration, 
and we simply computed $\Delta A_Z$
by locating the latter.

\section*{Acknowledgments}

I am grateful to H. Gies and O. Zanusso for precious discussions and advices.
I would like to thank T. R. Morris for correcting a mistake in a former version of this work.
It is a pleasure to stress that large parts of this research has been inspired by collaboration
with G. P. Vacca on related projects.
I acknowledge support by the DFG under grant GRK1523/2.

\appendix

\section{Large-field asymptotic expansion}
\label{sec:appendix}

We made use of the following truncated expansions
\be
V_{\text{as}}(\phi)\!&=&\!\left(1-\frac{\eta}{2}\right)\frac{\phi^2}{2}+A_V \phi^\frac{2d}{d+2-\eta}\nonumber\\
&&+\sum_{i=1}^{7}c_{V,i}(d,\eta,A_V) 
\phi^{e_{V,i}(d,\eta)}\nonumber\\ 
Z_{\text{as}}(\phi)^{-1}\!&=&\!A_Z \phi^\frac{4}{d-2+\eta}
+\sum_{i=1}^{9}c_{Z,i}(d,\eta,A_V,A_Z) \phi^{e_{Z,i}(d,\eta)}\nonumber
\ee
Here
\be
e_{V,1}(d,\eta)\!&=&\! 2\ \frac{d-2+\eta}{d+2-\eta}\nonumber\\
e_{V,2}(d,\eta)\!&=&\!0\nonumber\\
e_{V,3}(d,\eta)\!&=&\! -4+\frac{6d}{d+2-\eta}\nonumber\\
e_{V,4}(d,\eta)\!&=&\! -2+\frac{2d}{d+2-\eta}\nonumber\\ \nonumber
\ee
\be
e_{V,5}(d,\eta)\!&=&\! -6+\frac{8d}{d+2-\eta}\nonumber\\
e_{V,6}(d,\eta)\!&=&\! 4\ \frac{-2+\eta}{d+2-\eta}\nonumber\\
e_{V,7}(d,\eta)\!&=&\! 6\ \frac{-2+\eta}{d+2-\eta}\nonumber
\ee
and
\be
e_{Z,1}(d,\eta)\!&=&\! -2+\frac{2d}{d+2-\eta}
+\frac{4}{d-2+\eta}\nonumber\\
e_{Z,2}(d,\eta)\!&=&\! -4+\frac{4d}{d+2-\eta}
+\frac{4}{d-2+\eta}\nonumber\\
e_{Z,3}(d,\eta)\!&=&\! -2+\frac{4}{d-2+\eta}\nonumber\\
e_{Z,4}(d,\eta)\!&=&\! 6\ \frac{-2+\eta}{d+2-\eta}
+\frac{4}{d-2+\eta}\nonumber\\
e_{Z,5}(d,\eta)\!&=&\!  -4+\frac{2d}{d+2-\eta}
+\frac{4}{d-2+\eta}\nonumber\\
e_{Z,6}(d,\eta)\!&=&\! 8\ \frac{-2+\eta}{d+2-\eta}
+\frac{4}{d-2+\eta}\nonumber\\
e_{Z,7}(d,\eta)\!&=&\! -6+\frac{4d}{d+2-\eta}
+\frac{4}{d-2+\eta}\nonumber\\
e_{Z,8}(d,\eta)\!&=&\! 10\ \frac{-2+\eta}{d+2-\eta}
+\frac{4}{d-2+\eta}\nonumber \ .
\ee
Notice that the last order we took into account,
for $d=3$ and small enough $\eta>0$,
corresponds to the first negative power of $\phi$ inside $Z^{-1}$
(in the limiting case $\eta\to 0$ it gives $\phi^0$).
The first few coefficients in $V$ are
\be
c_{V,1}(d,\eta,A_V)\!&=&\! -\frac{4d^2 A_V^2}{(d+2-\eta)^2(-2+\eta)}\nonumber\\
c_{V,2}(d,\eta,A_V)\!&=&\! \frac{-2+\eta}{2d}\nonumber\\
c_{V,3}(d,\eta,A_V)\!&=&\! \frac{16 d^3 A_V^3(d-2+\eta)}{(d+2-\eta)^4(-2+\eta)^2}\nonumber\\
c_{V,4}(d,\eta,A_V)\!&=&\! -2d A_V \frac{d-2+\eta}{(d+2-\eta)^3}\nonumber\ .
\ee
Up to the computed order these depend only on $A_V$, while
the coefficient for $Z$
\be
c_{Z,1}(d,\eta,A_V,A_Z)\!&=&\! \frac{4d A_V A_Z}{(d+2-\eta)(-2+\eta)}\nonumber\\
c_{Z,2}(d,\eta,A_V,A_Z)\!&=&\! -\frac{32 d^2 A_V^2 A_Z}{(d+2-\eta)^3(-2+\eta)}\nonumber\\
c_{Z,3}(d,\eta,A_V,A_Z)\!&=&\! 4 A_Z\ \frac{d-6+\eta}{(d-2+\eta)^3}\nonumber\\
c_{Z,4}(d,\eta,A_V,A_Z)\!&=&\! \frac{64 d^3 A_V^3 A_Z (d-10+2\eta)}{(d+2-\eta)^5(-2+\eta)^2}\nonumber\ .
\ee
involve also $A_Z$. The remaining coefficients are too long to appear here.

%
%

\end{document}